\documentclass[
aps,%
11pt,%
final,%
notitlepage,%
oneside,%
onecolumn,%
nobibnotes,%
nofootinbib,%
superscriptaddress,%
noshowpacs,%
centertags]%
{revtex4}

\usepackage{multirow}
\renewcommand{\baselinestretch}{1.0}

\begin{document}

\selectlanguage{english}

\title{Galaxy Clusters and Their Outskirts: the ``Red Sequence'', Star Formation Rate, Stellar Mass}
\author{\firstname{F.~G.}~\surname{Kopylova}}

\author{\firstname{A.~I.}~\surname{Kopylov}}
\affiliation{\saoname}

\onecolumngrid
\begin{center}
{\scriptsize
Original Russian Text @ F.G.~Kopylova, A.I.~Kopylov,
published in Astrofizicheskii Byulleten, 2019,\\
Vol.74, No.4, pp.365-378}
\end{center}

\begin{abstract}
We study the outskirts ($R<3R_{200c}$) of 40 groups and
clusters of galaxies of the local Universe ($0.02<z<0.045$) with
300~km~s$^{-1}<\sigma<950$~km~s$^{-1}$).
Using the SDSS DR10 catalog data, we measured the stellar mass
of galaxy clusters in accordance with the previously determined
$K_s$-luminosity (2MASX data) and found their correlation in the form
$M_*/M_{\odot} \propto (L_K/L_{\odot})^{1.010\pm0.004}$ (\mbox{$M_K<-21\fm5$},
$R<R_{200c}$). We also found the dependence of the galaxy cluster
stellar mass on halo mass: \mbox{$M_*/M_{\odot} \propto
(M_{200c}/M_{\odot})^{0.77\pm0.01}$}. Our results show that the
fraction of galaxies with quenched star formation ($M_K<-21\fm5$)
is maximal in the central regions of the galaxy clusters and
equals, on the average, $0.81\pm0.02$; it decreases to
$0.44\pm0.02$ outside of the projected radius $R_{\rm sp}$
($2<R/R_{200c}<3$), which we found from the observed profile, but
still remains higher than that in the field by 27\%. The fraction
of early-type ``red sequence'' galaxies decreases from
$0.54\pm0.02$ in the center to $0.24\pm0.01$ beyond $R_{\rm sp}$,
reaching its field value.
\end{abstract}
\keywords{galaxies: clusters---galaxies: star
formation---galaxies: evolution}

\footnotetext{email: flera@sao.ru}
\maketitle

\section{INTRODUCTION}
It is known from observations that the high-density central
virialized regions of galaxy groups and clusters are populated by
massive red early-type galaxies without star formation (see,
e.g.,~\mbox{\cite{Dressler:Kopylova_n_en,Balogh:Kopylova_n_en,Kauffmann:Kopylova_n_en}}). It has also been
established that the properties of galaxies in a cluster are
related not only to the positions of the galaxies in it, but also
to the time of their entry into the system (see,
e.g.,~\cite{Hernandez:Kopylova_n_en,Muzzin:Kopylova_n_en,Mahajan:Kopylova_n_en}). Model orbits of galaxies
that fall within the gravitational potential of galaxy clusters
can be found in~\cite{Oman1:Kopylova_n_en,Oman2:Kopylova_n_en,Rhee:Kopylova_n_en}. The changes in the
properties of galaxies in groups and clusters with increasing
distance from the center have been investigated in many studies.
The main result is the discovery of a decreasing star formation
rate in galaxies in the direction of the cluster center
(e.g.,~\cite{Balogh1:Kopylova_n_en,Balogh2:Kopylova_n_en,Haines:Kopylova_n_en,Wetzel:Kopylova_n_en}). It was shown that
after a certain time, star formation fades in all galaxies of the
system~\cite{Oman2:Kopylova_n_en}. If galaxies with star formation are
occasionally found, then they have either entered the cluster
recently, or are not members at all but are merely projected onto
the region.

The authors of~\cite{Balogh2:Kopylova_n_en} found that the
decrease in the fraction of galaxies with star formation is
observed up to $5R_{200}$. To explain this effect, model
simulations were performed for the ($N$-body) star formation rate
change along the galaxy cluster radius. They showed that a
considerable fraction of galaxies beyond the virial cluster radius
have once been in their central regions---up to 50\% within
1--2~$R_{\rm vir}$~\cite{Gill:Kopylova_n_en,
Balogh2:Kopylova_n_en}.

The physical mechanisms that are present in the dense inner medium
of the clusters lead to a decrease in the amount of gas in the
galaxies and, in turn, to a decrease in the star formation rate.
The key mechanism leading to the venting of gas and a decrease in
the efficiency of the star formation process is assumed to be the
influence of AGN feedback in the
clusters~\cite{McNamara:Kopylova_n_en}. Over the course of several
Gyr, galaxies may lose the gaseous envelope even in the vicinity
of galaxy clusters~\cite{Bekki:Kopylova_n_en}. Or, if the density
of the cluster environment is sufficiently high, the cool gas disc
can be stripped off as a result of ram
pressure~\cite{Gunn:Kopylova_n_en,Quilis:Kopylova_n_en}.

Another hypothesis explains the quenching of star formation in the
outskirts of galaxy clusters by the fact that galaxies in the
groups that fall into the cluster have already undergone
pre-processing, having lost their gas as a result of the process
described above (see, e.g.,~\mbox{\cite{Zabludoff:Kopylova_n_en,
Wetzel:Kopylova_n_en, Haines:Kopylova_n_en,
Olave:Kopylova_n_en}}).

In order to establish the variation along the normalized
radius (up to $3R/R_{200}$) of the fraction of early-type ``red
sequence'' galaxies and the specific star formation rate, we used
in our study a sample of 40 galaxy groups and clusters with
registered X-ray emission, including systems of galaxies from the
Leo and Hercules supercluster regions, as well as several nearby
systems: clusters A\,1367, A\,1656 and eight groups of galaxies
considered earlier in~\cite{Kopylova1:Kopylova_n_en}. The sample
is formed in such a way that it would cover a maximum range of
radial velocity dispersions of the galaxy systems from
300~km~s$^{-1}$ to 950~km~s$^{-1}$ in the local Universe
(\mbox{$0.02<z<0.045$}). The results for six rich systems of
galaxies from the studied sample have already been published in
our paper~\cite{Kopylova0:Kopylova_n_en}.

In this work, we used data from the following catalogs:
SDSS\footnote{\url{\http://www.sdss.org}}~\cite{Ahn:Kopylova_n_en} (Sloan
Digital Sky Survey) Data Release~10, 2MASS (2MASX, Two-Micron
ALL-Sky Survey Extended Source
Catalog\footnote{\url{http://www.ipac.caltech.edu/2mass/releases/allsky/
}}~\cite{Jarrett:Kopylova_n_en}), and
NED\footnote{\url{\http://nedwww.ipac.caltech.edu}} (NASA
Extragalactic Database).

The paper is organized as follows. In Section 2 we describe the
procedure of selecting the outskirts of the galaxy groups and
clusters (in units of radii $R_{200c}$ and $R_{\rm sp}$). In
Section~3 we present the constructed color--magnitude sequence for
early-type galaxies in the systems of galaxies and determine the
fraction of early-type galaxies. Section~4 describes our
measurements of the total stellar mass of the system of galaxies
within the radius $R_{200c}$ and compares the obtained stellar
mass with the dynamical mass determined earlier. In Section~5 we
give the distributions of the specific star formation rate of the
clusters with fixed galaxy stellar mass and without such. We show
variations of the fraction of quenched galaxies (QG) along the
normalized radius of the systems with fixed galaxy stellar mass
and without such, and for comparison we estimate the fractions of
field galaxies with quenched star formation. In the Conclusions we
list the main results. In this work we used the following
cosmological parameters: $\Omega_m=0.3$,
\mbox{$\Omega_{\Lambda}=0.7$},
\mbox{$H_0=70$}~km~s$^{-1}$\,Mpc$^{-1}$.

\section{THE NEAREST OUTSKIRTS OF GALAXY GROUPS AND CLUSTERS}

We determined the dynamical characteristics of the studied systems
of galaxies: radial velocity, radial velocity dispersion, cluster
mass (halo) within the  radius $R_{200c}$ (in projection).
The $R_{200c}$ radius is the radius of a sphere within which the
density in the system exceeds the critical density of the Universe
by a factor of 200. It is determined by the radial velocity
dispersion of the system galaxies~\cite{Kopylova0:Kopylova_n_en}. Model
simulations often use another radius, $R_{200m}$ --- the radius of a
sphere within which the density in the system is over 200 times
the average density of the Universe.

The relations between the obtained masses $M_{200c}$ and
$K_s$-luminosity (2MASX), as well as the mass--luminosity ratio
and the dynamical parameters (based on SDSS\,DR7, DR8 data) for
the galaxy groups and clusters under study are presented in our
earlier
papers~\mbox{\cite{Kopylova2:Kopylova_n_en,Kopylova3:Kopylova_n_en,Kopylov:Kopylova_n_en}}.

\begin{figure*}[]
\setcaptionmargin{5mm} \onelinecaptionstrue
\includegraphics[scale=0.53,angle=0]{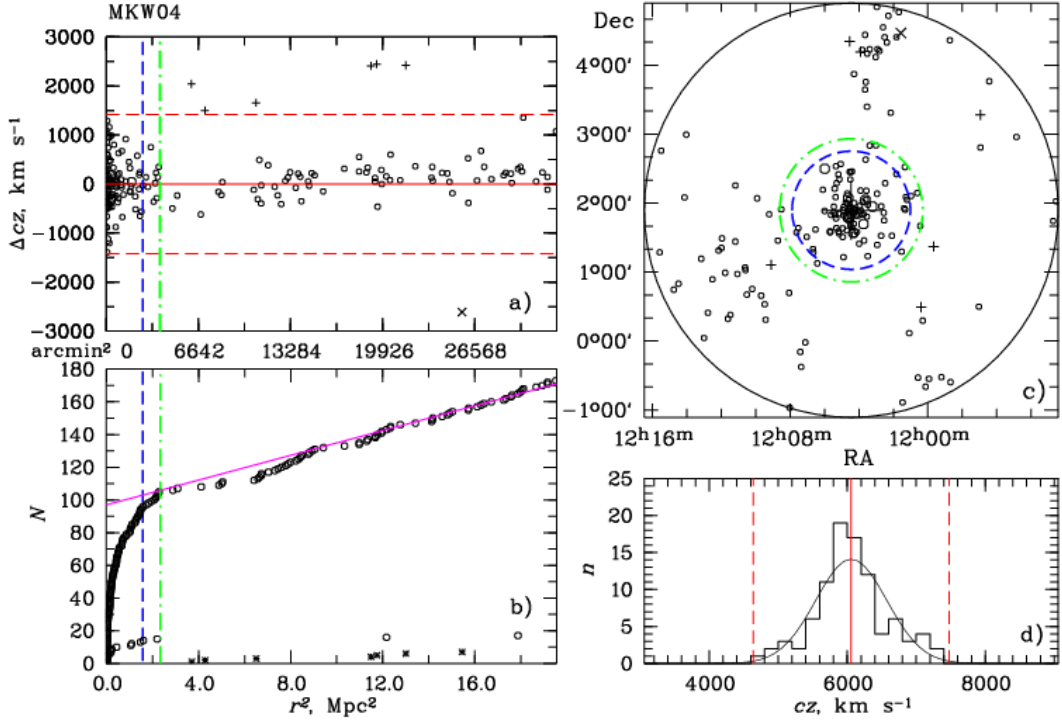}
\captionstyle{normal} \caption{ Distribution of galaxies in the
cluster MKW\,04. Panel (a) shows the deviation of galaxy radial
velocities from
the average radial velocity of the cluster, determined from the
galaxies within the radius $R_{200c}$. The horizontal dashed lines
correspond to deviations $\pm2.7\sigma$, the vertical dashed line
shows the radius $R_{200c}$, and the dashed and dotted line
represents radius $R_{\rm sp}$. The larger circles denote galaxies
brighter than $M_K^* = -24^{\rm m}$, the straight crosses mark the
background galaxies, the diagonal crosses show the foreground
galaxies. Panel (b) shows the integrated distribution of the
total number of
galaxies (upper curve) as a function of squared distance to the
group center. The lower points correspond to early-type galaxies
brighter than $M_K<-21\fm5$. The circles correspond to the
galaxies marked by circles in panel~(a), and the stars show the
background galaxies.
Panel (c) shows the sky distribution in equatorial coordinates for
galaxies shown in panel~(a) (the designations are the same).
The region under study is
restricted by a circle with the radius equal to $3.5\,R_{200c}$
(solid line). The center of the cluster is marked by a large
cross. Panel (d) shows the radial velocity distribution for all the galaxies
within the radius $R_{200c}$ (the Gaussian (solid line)
corresponding to the velocity dispersion of the group is shown for
the cluster members). The solid vertical line shows the average
radial velocity of the group, and the dashed lines correspond to
deviations of $\pm2.7\sigma$.} \label{clusMKW04:Kopylova_n_en}
\end{figure*}

In our study, it is important to isolate the nearest outskirts of
the galaxy systems: in addition to the radius $R_{200c}$, we must
determine the ``splashback'' radius $R_{\rm sp}$ \mbox {($R_{\rm
sp}$>$R_{200c}$)}. We analyze the distributions which characterize
in detail the structure and kinematics of each system, and present
as figures the following:
\begin{list}{}{
\setlength\leftmargin{2mm} \setlength\topsep{2mm}
\setlength\parsep{0mm} \setlength\itemsep{2mm} }
 \item 1)~the deviation of the radial velocities of cluster member galaxies and background galaxies from the average radial velocity of the cluster or group as a function of squared radius (the distance to the cluster center);
 \item 2)~the integrated distribution of the total number of galaxies as a function of squared radius;
 \item 3)~the position of the galaxies on the sky plane in equatorial coordinates;
\item 4)~the histogram of the distribution of the radial
velocities of all galaxies within the radius $R_{200c}$. As an
example, we show in Figs.~\ref{clusMKW04:Kopylova_n_en}a--\ref{clusMKW04:Kopylova_n_en}d such
graphs for cluster MKW\,04.
\end{list}

Let us consider Fig.~\ref{clusMKW04:Kopylova_n_en}b, where we present the
projected cluster profile. As the distance from the cluster center
increases, we observe first a sharp increase in the number of
galaxies, and then, beyond the limits of the virialized
region~\cite{Kopylov:Kopylova_n_en}, its linear growth (shown in the figure by a
straight line). The radius of the virialized region $R_{200c}$ and
the radius $R_{\rm sp}$ are shown by the dashed and dashed and
dotted lines. Note that $R_{\rm sp}$ corresponds to the distance
from the center where the sharp increase in the number of cluster
members changes to linear. In the bottom of this figure we also
show the distribution of early-type galaxies brighter than
$M_K=-21\fm5$, which we used to refine the radius. Such galaxies
are usually located in the central virialized regions of galaxy
systems. We assume that the radius $R_{\rm sp}$ corresponds to the
apocenter of the orbits for galaxies that entered the virialized
region of the cluster and then flew out to come back again. In
other words, it separates the galaxies which enter the cluster for
the first time from those that are already cluster members.

According to model ($N$-body) simulations~\cite{Mamon:Kopylova_n_en}, the
maximum distance of the galaxies from the clusters does not
exceed 1--2.5~$R_{\rm vir}$. Model simulations for the dark halo
profile of the clusters~\cite{More1:Kopylova_n_en} have shown that for rapidly
accreting halos, the galaxy cluster radius where a sharp increase
occurs in the slope of the halo surface density distribution is
0.8--1.0~$R_{200m}$, and the $R_{\rm sp}\sim
1.5~R_{200m}$ for slowly accreting halos.

Based on observation of nearby groups of the Local Universe,
Tully ~\cite{Tully:Kopylova_n_en} derived the radius \mbox
{$R_{\rm sp} \approx 1.33 R_{200c}$}, called it ''the second
turnaround radius``. For distant galaxies, the $R_{\rm sp}$ radii
were found based on the SDSS catalog data from the galaxy surface
density profile and from the mass profile determined by weak
gravitational
lensing~\cite{More2:Kopylova_n_en,Chang:Kopylova_n_en}. The radius
$R_{\rm sp}$ obtained by us from the integrated distribution of the
number of galaxies as a function of squared distance from the
cluster center is equal, on average, to $(1.54\pm0.06)~R_{200c}$
for 40 studied systems, which amounts to  \mbox
{$0.96\pm0.04~R_{200m}$} (considering that \mbox {$4R_{200c}
\approx 2.5R_{200m}$}). This value is larger than the estimate
in~\cite{Tully:Kopylova_n_en}, but agrees with the results derived
from model simulations, for example~\cite{More1:Kopylova_n_en}.

\renewcommand\baselinestretch{0.90}
\begin{table*}[]
\setcaptionmargin{0mm} \onelinecaptionstrue \captionstyle{normal}
\caption{Physical properties of galaxy clusters} \label{data1:Kopylova_n_en}
\medskip
\begin{tabular}{l|c|c|c|c|c|c|c|r@{$\,\pm\,$}l|c}
\hline
\multicolumn{1}{c|}{\multirow{2}{*}{Cluster}} & $\sigma$,    & $R_{200c}$, & $R_{200m}$, &  $R_{\rm sp}$, & \multirow{2}{*}{$N_z$} & \multirow{2}{*}{$z_h$} &  $M_{200c}$,        & \multicolumn{2}{c|}{$L_{K,200c}$,}       & $M_{*,200c}$, \\
                                              & km~s$^{-1}$ & Mpc         & Mpc         & Mpc            &                        &                        & $10^{14}~M_{\odot}$ & \multicolumn{2}{c|}{$10^{12}~M_{\odot}$} & $10^{12}~M_{\odot}$\\
\hline
\multicolumn{1}{c|}{(1)}&(2)&(3)&(4)&(5)&(6)&(7)&(8)&\multicolumn{2}{c|}{(9)}&(10)\\
\hline
A\,2147     & $853\pm46$ & 2.076 & 3.32 & 3.64  & 344 & 0.036179 & $10.57\pm1.71$ &$12.68$ & $0.6$ & 11.34 \\
A\,2063     & $753\pm62$ & 1.833 & 2.93 & 2.68  & 146 & 0.034664 & $7.28\pm1.80$  &$5.97 $ & $0.4$ &  5.78 \\
A\,1367     & $749\pm47$ & 1.835 & 2.94 & 2.78  & 249 & 0.021743 & $7.21\pm1.46$  &$7.62 $ & $0.2$ &  7.25 \\
A\,2199     & $746\pm44$ & 1.820 & 2.91 & 3.61  & 288 & 0.030458 & $7.09\pm1.25$  &$9.88 $ & $0.3$ &  8.84 \\
A\,1185     & $676\pm52$ & 1.692 & 2.70 & 2.97  & 177 & 0.032883 & $5.70\pm1.28$  &$6.68 $ & $0.2$ &  6.16 \\
MKW\,03s    & $608\pm67$ & 1.474 & 2.36 & 1.95  &  82 & 0.044953 & $3.81\pm1.26$  &$4.56 $ & $0.4$ &  4.60 \\
NGC\,6338   & $552\pm61$ & 1.348 & 2.16 & 2.12  &  83 & 0.029342 & $2.87\pm0.94$  &$2.75 $ & $0.1$ &  2.52 \\
NGC\,6107   & $546\pm55$ & 1.332 & 2.13 & 1.90  &  99 & 0.031093 & $2.78\pm0.84$  &$3.45 $ & $0.1$ &  3.62 \\
RXJ\,1722   & $524\pm86$ & 1.269 & 2.03 & 1.64  &  37 & 0.046580 & $2.44\pm1.20$  &$3.58 $ & $0.6$ &  3.92 \\
MKW\,04     & $515\pm54$ & 1.263 & 2.02 & 1.53  &  91 & 0.020208 & $2.34\pm0.74$  &$2.50 $ & $0.1$ &  2.57 \\
UGC\,04991  & $515\pm68$ & 1.256 & 2.01 & 1.72  &  58 & 0.031958 & $2.33\pm0.92$  &$2.26 $ & $0.2$ &  2.06 \\
A\,1983     & $460\pm47$ & 1.115 & 1.78 & 1.67  &  97 & 0.044803 & $1.65\pm0.51$  &$4.86 $ & $0.6$ &  4.50 \\
MKW\,08     & $450\pm44$ & 1.100 & 1.76 & 1.73  & 103 & 0.026906 & $1.56\pm0.46$  &$3.08 $ & $0.1$ &  3.10 \\
NGC\,5098   & $445\pm58$ & 1.083 & 1.73 & 1.73  &  58 & 0.036812 & $1.50\pm0.59$  &$2.81 $ & $0.1$ &  3.21 \\
RBS\,858    & $445\pm63$ & 1.081 & 1.73 & 1.64  &  50 & 0.039586 & $1.51\pm0.64$  &$2.37 $ & $0.2$ &  2.29 \\
NGC\,2795   & $425\pm61$ & 1.039 & 1.66 & 1.33  &  49 & 0.028887 & $1.31\pm0.56$  &$2.08 $ & $0.02$&  1.80 \\
MKW\,04s    & $423\pm56$ & 1.033 & 1.65 & 1.53  &  56 & 0.027928 & $1.29\pm0.51$  &$1.83 $ & $0.1$ &  1.78 \\
VV\,196     & $412\pm69$ & 1.003 & 1.60 & 1.16  &  36 & 0.035289 & $1.19\pm0.60$  &$1.29 $ & $0.05$&  1.16 \\
RXJ\,1033   & $411\pm65$ & 0.996 & 1.59 & 1.48  &  40 & 0.045170 & $1.18\pm0.56$  &$2.23 $ & $0.2$ &  2.21 \\
AWM\,1      & $402\pm58$ & 0.982 & 1.57 & 1.15  &  48 & 0.028652 & $1.11\pm0.48$  &$2.29 $ & $0.1$ &  1.92 \\
AWM\,4      & $380\pm62$ & 0.927 & 1.48 & 1.34  &  37 & 0.031827 & $0.94\pm0.46$  &$1.49 $ & $0.1$ &  1.40 \\
NGC\,7237   & $376\pm52$ & 0.919 & 1.47 & 1.47  &  52 & 0.026102 & $0.91\pm0.38$  &$1.53 $ & $0.1$ &  1.52 \\
NGC\,3158   & $375\pm61$ & 0.918 & 1.47 & 1.24  &  38 & 0.022630 & $0.90\pm0.44$  &$1.74 $ & $0.1$ &  1.70 \\
RXC\,J1511  & $374\pm76$ & 0.909 & 1.45 & 1.22  &  24 & 0.038990 & $0.89\pm0.54$  &$1.22 $ & $0.1$ &  0.98 \\
NGC\,5171   & $371\pm52$ & 0.908 & 1.47 & 1.47  &  51 & 0.023000 & $0.88\pm0.37$  &$1.46 $ & $0.1$ &  1.23 \\
NGC\,3119   & $355\pm59$ & 0.867 & 1.39 & 1.20  &  36 & 0.029657 & $0.76\pm0.38$  &$1.50 $ & $0.1$ &  1.71 \\
A\,1228B    & $347\pm66$ & 0.842 & 1.35 & 1.28  &  28 & 0.042892 & $0.71\pm0.40$  &$1.74 $ & $0.1$ &  1.48 \\
A\,2162     & $346\pm61$ & 0.844 & 1.35 & 1.12  &  32 & 0.032147 & $0.71\pm0.37$  &$1.44 $ & $0.1$ &  1.58 \\
NGC\,2783   & $346\pm58$ & 0.848 & 1.36 & 1.02  &  35 & 0.022151 & $0.71\pm0.38$  &$1.07 $ & $0.1$ &  1.04 \\
A\,1177     & $337\pm66$ & 0.822 & 1.32 & 1.14  &  26 & 0.032159 & $0.65\pm0.35$  &$1.20 $ & $0.1$ &  1.14 \\
NGC\,6098   & $335\pm77$ & 0.817 & 1.31 & 1.14  &  19 & 0.030936 & $0.64\pm0.44$  &$0.87 $ & $0.1$ &  0.95 \\
UGC\,07115  & $334\pm46$ & 0.818 & 1.31 & 1.06  &  53 & 0.022199 & $0.64\pm0.26$  &$1.29 $ & $0.1$ &  1.18 \\
NGC\,2832   & $331\pm43$ & 0.810 & 1.30 & 1.20  &  60 & 0.023044 & $0.62\pm0.24$  &$1.59 $ & $0.1$ &  1.03 \\
NGC\,5627   & $314\pm55$ & 0.768 & 1.23 & 1.10  &  33 & 0.026682 & $0.53\pm0.28$  &$1.34 $ & $0.1$ &  1.18 \\
\hline
\end{tabular}
\end{table*}
\renewcommand\baselinestretch{1.0}

The physical properties of the groups and clusters of galaxies
considered in this work for the region of radius $R_{200c}$ are
presented in the columns of Table~1:  (1)~name of the galaxy
cluster; (2)~radial velocity dispersion with the cosmological
correction \mbox {$(1+z)^{-1}$}; \mbox {(3)--(5)}~radii
$R_{200c}$, $R_{200m}$, and $R_{\rm sp}$ in Mpc; (6)~number of
galaxies for which the redshift and radial velocity dispersion are
known; (7)~measured redshift; (8)~halo mass $M_{200c}$ determined
from $\sigma$, with an error that corresponds to the $\sigma$
measurement error; (9)~$L_{K,200c}$--luminosity with the error
determined in the same way; (10)~stellar mass $M_{*,200c}$ with
the error approximately the same as that for the luminosity.
stellar mass determination is described in detail in Section~4.

\subsection{Comments on Some Galaxy Clusters}

In the galaxy cluster pair A\,2199\,+\,A\,2197, A\,2199 is a
richer system with a higher X-ray luminosity~\cite{Kopylova3:Kopylova_n_en}.
Both clusters are projected close to each other on the celestial
sphere and have practically identical radial velocities. A\,2197
is separated by approximately 79~arc minutes in $\delta$ from
A\,2199 and, according to our measurements, is located within the
$R_{\rm sp}$ radius of cluster A\,2199, i.e., has common outskirts
with A\,2199. Therefore, in our work the cluster A\,2197 is not
considered individually.

In the galaxy cluster pair  A\,2147\,+\,A\,2151, A\,2147 is also a
richer system with a higher X-ray luminosity~\cite{Kopylova3:Kopylova_n_en}.
These clusters are located close to each other in celestial sphere
projection and have practically identical radial velocities.
However, cluster A\,2151 is located further, approximately 105 arc
minutes by $\delta$ from A\,2147. Although according to our
measurements A\,2151 does not coincide with the $R_{\rm sp}$
radius of A\,2147, we also do not consider it separately, since
both clusters have common outskirts.

\section{THE ``RED SEQUENCE'' OF GALAXY GROUPS AND CLUSTERS}

Early type galaxies are the brightest, and therefore, the most
massive members of galaxy clusters. There are located for the most
part in the central virialized regions (see,
e.g.,~\cite{Raichoor:Kopylova_n_en}). Thus, according to our estimates, in the
clusters of the Hercules supercluster they account for about
60--70\% of galaxies brighter than $M_K=-23\fm3$~\cite{Kopylova1:Kopylova_n_en}.
The fact that early type galaxies are the brightest objects in the
clusters is also confirmed by the luminosity functions for early-
and late-type galaxies presented in the paper.

\begin{figure*}[]
\setcaptionmargin{5mm} \onelinecaptionstrue
\includegraphics[scale=0.55,angle=0]{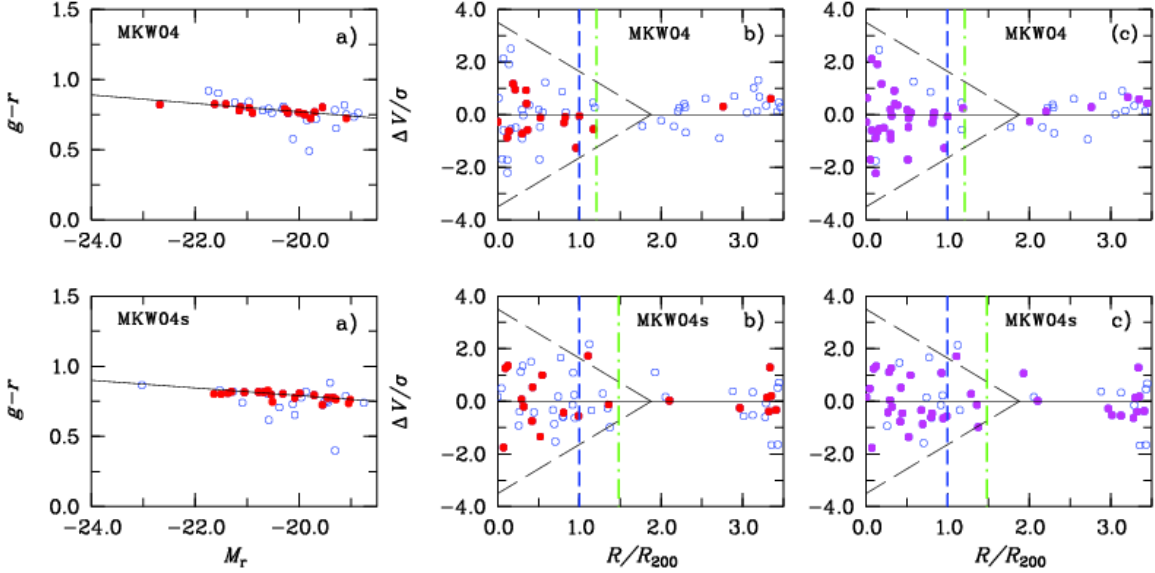}
\captionstyle{normal} \caption{Color--magnitude relation
$(g-r)_o$ vs. $M_r$ for galaxy clusters
MKW\,04 and MKW\,04s (panels a). The filled circles show
early type galaxies of the ``red sequence''. The panels (b)
shows phase-space diagrams --- velocity of galaxy clusters
as a function of radius. The velocity is determined as
the ratio of the difference between the radial velocities of the
galaxies and the average radial velocity of the cluster to radial
velocity dispersion. The $R/R_{200c}$ radius is the distance of
the galaxy from the center, normalized to the $R_{200c}$ radius.
Early-type RS-galaxies are shown by filled circles. The dashed and
dashed-and-dotted lines correspond to the $R_{200c}$ and $R_{\rm
sp}$ radii. The galaxies are selected within
$\pm2.7\sigma$~\cite{Mamon1:Kopylova_n_en}. The panels (c)  shows the same
phase-space diagrams, but with the filled circles showing galaxies with
quenched star formation ($\log sSFR [{\rm Gyr^{-1}}] < -1.75$).
The tilted model black dashed lines to separate the virialized
galaxies  from those that fall into the cluster for the first time
\cite{Barsanti:Kopylova_n_en}.}
\label{3figs:Kopylova_n_en}
\end{figure*}

One of the fundamental relations for early type galaxies is the
dependence of color on magnitude, which has a slight scatter for
groups and clusters of galaxies and forms the so-called ``red
sequence''~(RS). The RS is related to the physical properties of
the systems: its shape characterizes the mass--metallicity
relation, and its scatter shows the galaxy age variations (see,
e.g.,~\cite{Gladders:Kopylova_n_en}). The early-type galaxy RS is
often used in search activities. It can usually be used to find
galaxy clusters in the
near~\cite{Garilli:Kopylova_n_en,Scodeggio:Kopylova_n_en,Lopez-Cruz:Kopylova_n_en}
and far Universe~\mbox{\cite{Aragon-Salamanca:Kopylova_n_en,
Stanford:Kopylova_n_en, van Dokkum:Kopylova_n_en,
Blakeslee:Kopylova_n_en, Andreon2:Kopylova_n_en}}. In this work,
we used early type RS galaxies brighter than \mbox{$M_K=-21\fm5$}
in order to study in more detail the outskirts of the galaxy
clusters under investigation ($R<3R_{200c}$).

We used the following criteria in our selection of galaxies:
$fracDeV\geq 0.8$, where $fracDeV$ is a galaxy parameter which
represents the input of the bulge into the total luminosity of the
galaxy, based on the SDSS $r$-band catalog data;
\mbox{$c=r_{90}/r_{50}\geq 2.6$}, where $c$ is the concentration
index, equal to the ratio of radii that limit $90\%$ and $50\%$ of
the Petrosian fluxes. Additionally, we imposed a color restriction
for $u-r$: $\Delta (u-r)_o > -0.2$, which follows from the obtained
dependence of the $u-r$ color on the Petrosian magnitude for the
galaxy clusters, e.g., the Hercules supercluster: $u-r = -0.078
\times r_{\rm Pet}+3.81$ with \mbox {$2\sigma = -0.2$}.

For the studied systems, we constructed a series of figures
similar to the ones in Fig.~\ref{3figs:Kopylova_n_en} for the galaxy clusters
MKW\,04 and MKW\,04s. Figs.~\ref{3figs:Kopylova_n_en}a show the
color--magnitude diagrams with the ``red sequence''. Early-type
galaxies on the RS that satisfy the $-0.075<g-r<0.075$ condition
are shown by filled circles. Figs.~\ref{3figs:Kopylova_n_en}(b) and (c) show
the projected phase-space diagrams for the galaxy clusters MKW\,04
and MKW\,04s, where $\Delta V/\sigma$ is the ratio of the
difference of the galaxy radial velocity and the average radial
velocity of the cluster to the radial velocity dispersion of the
cluster, and $R/R_{200c}$ is the distance of the galaxy from the
selected cluster center normalized to radius $R_{200c}$.

The projected phase-space diagram shows the dependence between the
distance of the galaxy from the selected center and its usually
normalized radial velocity. Based on the positions of the galaxies
on the diagram one can investigate, for example, the variations of
star formation in the clusters, and also make conclusions about
galaxies belonging to the ``old'' population of the galaxy cluster
or to galaxies that have fallen into it only recently, of the
those that have already flew out of the virialized region
(``splashback''-galaxies) \mbox {\cite{Mahajan:Kopylova_n_en,
Hernandez:Kopylova_n_en, Muzzin:Kopylova_n_en,
Oman1:Kopylova_n_en, Oman2:Kopylova_n_en, Rhee:Kopylova_n_en}}.
Analyzing the position of a galaxy on the phase-space diagram, one
can conveniently study the history of its accretion into the group
or cluster.

As the galaxy system center we used the coordinates of the
brightest galaxy which is usually located not far from the center
of the X-ray emitting region. If there are several bright galaxies
or the bright galaxy is shifted from the galaxy concentration
center, then the x-ray emitting center or the galaxy centroid are
usually taken as the center. Fig.~\ref{3figs:Kopylova_n_en} shows only the
galaxies brighter than $M_K = -21\fm5$. In panels (b)\mbox{--}(c)
we used the tilted dashed lines to separate the mostly virialized
galaxies from those that fall into the cluster for the first time
(model simulations were taken from~\cite{Barsanti:Kopylova_n_en}).

Note that the main bulk of early-type RS galaxies is located in
the central regions of the galaxy clusters, within the derived
radius $R_{\rm sp}$. At the same time, a small portion of galaxies
may also be located at a distance of $3R_{200c}$, in accordance
with model simulations, e.g.,~\cite{Haines:Kopylova_n_en}. We found that the
fraction of early-type RS galaxies decreases, based on the
all-system average, by a factor of two: from $0.54\pm0.02$ in the
central regions to $0.24\pm0.02$ in  \mbox
{$2R/R_{200c}<R<3R/R_{200c}$}. Beyond the $R_{\rm sp}$ radii the
fraction of early type galaxies $frac_E$ also amounts to \mbox
{$0.24\pm0.01$}, on average. The results of $frac_E$ determination
along the normalized radius of the studied system are presented in
Table~\ref{data2:Kopylova_n_en}. Column~(1) gives the name of the system, and
the remaining columns give the variation ranges for radii
($R/R_{200c}$ and $R_{\rm sp}$).

\renewcommand\baselinestretch{0.90}
\begin{table*}[]
\setcaptionmargin{0mm} \onelinecaptionstrue \captionstyle{normal}
\caption{Variations of the fraction of early-type galaxies along
the radius} \label{data2:Kopylova_n_en}
\medskip
\begin{tabular}{l|c|c|c|c|c|c}
\hline
\multicolumn{1}{c|}{Cluster} & (0--0.25)$R/R_{200c}$ & (0--1)$R/R_{200c}$ & (1--2)$R/R_{200c}$ &  (2--3)$R/R_{200c}$ & (0--1)$R_{\rm sp}$ & 1$R_{\rm sp}$--3$R/R_{200c}$\\
\hline
\multicolumn{1}{c|}{(1)}&(2)&(3)&(4)&(5)&(6)&(7)\\
\hline
A\,2147      &  $0.45\pm0.11$ &$0.37\pm0.04$  & $0.28\pm0.04$   & $0.31\pm0.04$ & $0.34\pm0.03$ & $0.30\pm0.03$\\
A\,2063      &  $0.62\pm0.14$ &$0.50\pm0.08$  & $0.23\pm0.07$   & $0.28\pm0.06$ & $0.46\pm0.07$ & $0.26\pm0.05$\\
A\,1367      &  $0.58\pm0.14$ &$0.47\pm0.07$  & $0.32\pm0.08$   & $0.15\pm0.06$ & $0.44\pm0.06$ & $0.21\pm0.06$\\
A\,2199      &  $0.69\pm0.15$ &$0.46\pm0.06$  & $0.33\pm0.06$   & $0.18\pm0.04$ & $0.40\pm0.04$ & $0.18\pm0.04$\\
A\,1185      &  $0.49\pm0.14$ &$0.30\pm0.05$  & $0.33\pm0.07$   & $0.23\pm0.07$ & $0.32\pm0.05$ & $0.25\pm0.06$\\
MKW\,03s     &  $0.62\pm0.22$ &$0.49\pm0.09$  & $0.32\pm0.11$   & $0.21\pm0.10$ & $0.47\pm0.08$ & $0.22\pm0.08$\\
NGC\,6338    &  $0.47\pm0.20$ &$0.29\pm0.09$  & $0.26\pm0.12$   & $0.25\pm0.09$ & $0.30\pm0.08$ & $0.23\pm0.08$\\
NGC\,6107    &  $0.47\pm0.19$ &$0.43\pm0.09$  & $0.45\pm0.14$   & $0.14\pm0.06$ & $0.44\pm0.09$ & $0.21\pm0.07$\\
RXC\,J1722   &  $0.50\pm0.23$ &$0.44\pm0.12$  & $0.24\pm0.13$   & $0.27\pm0.15$ & $0.42\pm0.10$ & $0.25\pm0.11$\\
MKW\,04      &  $0.33\pm0.15$ &$0.41\pm0.12$  & $0.25\pm0.28$   & $0.09\pm0.09$ & $0.40\pm0.11$ & $0.08\pm0.09$\\
UGC\,04991   &  $0.39\pm0.17$ &$0.35\pm0.10$  & $0.38\pm0.18$   & $0.22\pm0.12$ & $0.39\pm0.10$ & $0.21\pm0.09$\\
A\,1983      &  $0.54\pm0.19$ &$0.52\pm0.09$  & $0.26\pm0.08$   & $0.24\pm0.12$ & $0.45\pm0.07$ & $0.22\pm0.08$\\
MKW\,08      &  $0.54\pm0.28$ &$0.49\pm0.12$  & $0.26\pm0.10$   & $0.29\pm0.11$ & $0.43\pm0.09$ & $0.28\pm0.09$\\
NGC\,5098    &  $0.54\pm0.25$ &$0.46\pm0.12$  & $0.19\pm0.08$   & $0.39\pm0.14$ & $0.35\pm0.08$ & $0.37\pm0.12$\\
RBS\,858     &  $0.62\pm0.28$ &$0.46\pm0.12$  & $0.50\pm0.22$   & $0.12\pm0.13$ & $0.48\pm0.12$ & $0.29\pm0.16$\\
NGC\,2795    &  $0.46\pm0.24$ &$0.34\pm0.13$  & $0.21\pm0.14$   & $0.40\pm0.17$ & $0.40\pm0.12$ & $0.28\pm0.11$\\
MKW\,04s     &  $0.60\pm0.44$ &$0.37\pm0.13$  & $0.25\pm0.36$   & $0.50\pm0.43$ & $0.35\pm0.11$ & $0.40\pm0.33$\\
VV\,196      &  $0.75\pm0.33$ &$0.48\pm0.17$  & $0.06\pm0.06$   & $0.36\pm0.19$ & $0.41\pm0.14$ & $0.23\pm0.10$\\
RXC\,J1033   &  $0.67\pm0.30$ &$0.38\pm0.11$  & $0.26\pm0.11$   & $0.13\pm0.10$ & $0.34\pm0.08$ & $0.20\pm0.11$\\
AWM\,1       &  $0.33\pm0.27$ &$0.39\pm0.14$  & $0.33\pm0.17$   & $0.42\pm0.22$ & $0.40\pm0.14$ & $0.36\pm0.14$\\
AWM\,4       &  $0.70\pm0.34$ &$0.43\pm0.08$  & $0.10\pm0.10$   & $0.14\pm0.06$ & $0.39\pm0.07$ & $0.05\pm0.05$\\
NGC\,7237    &  $0.46\pm0.24$ &$0.39\pm0.12$  & $0.36\pm0.19$   & $0.25\pm0.28$ & $0.39\pm0.10$ & $0.20\pm0.22$\\
NGC\,3158    &  $0.50\pm0.25$ &$0.48\pm0.10$  & $0.14\pm0.15$   & --            & $0.46\pm0.09$ & --           \\
RXC\,J1511   &  $0.57\pm0.21$ &$0.58\pm0.12$  & $0.27\pm0.18$   & $0.36\pm0.21$ & $0.54\pm0.10$ & $0.33\pm0.16$\\
NGC\,5171    &  $0.31\pm0.09$ &$0.40\pm0.07$  & $0.33\pm0.17$   & $0.22\pm0.17$ & $0.39\pm0.06$ & $0.20\pm0.15$\\
NGC\,3119    &  $0.67\pm0.35$ &$0.28\pm0.06$  & $0.11\pm0.08$   & --            & $0.24\pm0.04$ & $0.07\pm0.07$\\
A\,1228B     &  $0.50\pm0.25$ &$0.35\pm0.14$  & $0.28\pm0.14$   & $0.50\pm0.27$ & $0.33\pm0.11$ & $0.40\pm0.19$\\
A\,2162      &  $0.43\pm0.30$ &$0.43\pm0.17$  & $0.36\pm0.19$   & $0.38\pm0.25$ & $0.45\pm0.15$ & $0.29\pm0.16$\\
NGC\,2783    &  $0.29\pm0.12$ &$0.27\pm0.07$  & $0.40\pm0.33$   & $0.60\pm0.44$ & $0.29\pm0.07$ & $0.50\pm0.31$\\
A\,1177      &  $0.71\pm0.27$ &$0.55\pm0.12$  & $0.50\pm0.27$   & $0.50\pm0.43$ & $0.52\pm0.10$ & $0.57\pm0.36$\\
NGC\,6098    &  $0.67\pm0.61$ &$0.30\pm0.20$  & $0.20\pm0.22$   & --            & $0.21\pm0.14$ & $0.50\pm0.61$\\
UGC\,07115   &  $0.40\pm0.33$ &$0.57\pm0.18$  & $0.29\pm0.23$   & --            & $0.54\pm0.16$ & --           \\
NGC\,2832    &  $0.50\pm0.35$ &$0.39\pm0.08$  & $0.47\pm0.19$   & $0.60\pm0.31$ & $0.42\pm0.07$ & $0.56\pm0.23$\\
NGC\,5627    &  $0.29\pm0.23$ &$0.21\pm0.12$  & $0.40\pm0.24$   & $0.17\pm0.18$ & $0.27\pm0.11$ & $0.22\pm0.17$\\
\hline
Total $N=40$ &  $0.54\pm0.04$ &$0.44\pm0.02$  & $0.31\pm0.01$   & $0.24\pm0.02$ & $0.42\pm0.02$ & $0.24\pm0.01$\\
\hline
\end{tabular}
\end{table*}
\renewcommand\baselinestretch{1.0}

\begin{figure}[]
\setcaptionmargin{5mm} \onelinecaptionstrue
\includegraphics[scale=0.35,angle=0]{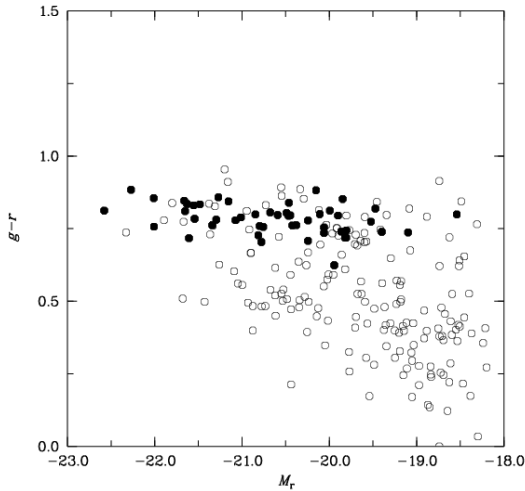}
\captionstyle{normal} \caption{Color--absolute magnitude
dependence for the field galaxies in the region with the center
$14\fh5$, $35\degr$ and a 300$\arcmin$ radius ($M_K<-21\fm5$,
$0.030<z<0.045$). Early type galaxies are shown by filled
circles.} \label{field:Kopylova_n_en}
\end{figure}

In our paper~\cite{Kopylova0:Kopylova_n_en} we considered two
regions between the superclusters Hercules and Leo, free of galaxy
clusters (practically field), and derived that the two-field
average for the fraction of early-type galaxies brighter than
$M_K=-21\fm5$ is equal to \mbox{$0.24\pm0.01$}. Unlike the cluster
galaxies, early type galaxies in these regions (field galaxies)
are not closely positioned on the RS, although they also
demonstrate a color--magnitude
dependence~(Fig.~\ref{field:Kopylova_n_en}).

As we derived above, for the nearest outskirts of galaxy clusters
\mbox{$2<R/R_{200c}<3$} and $R>R_{\rm sp}$ the average fractions
are $0.24\pm0.02$ and \mbox{$0.24\pm0.01$} (Table~\ref{data2:Kopylova_n_en}),
i.e. they are not different from $\langle frac_E \rangle$ for
field galaxies. Within the radius $1<R/R_{200c}<2$ the average
fraction of early type galaxies is higher than in the field, and
is equal to $0.31\pm0.01$. Thus, the main portion of early type
galaxies in the cluster lies within the limits of the radius
$R_{\rm sp}$ determined from observations: some of them are
already virialized, others (according to, e.g., model simulations
in~\cite{Rhee:Kopylova_n_en}) have left the virialized regions and, having
reached the apocenter of the orbit $R_{\rm sp}$ that we found,
will return into it again.

\section{RELATION BETWEEN THE STELLAR MASS AND THE TOTAL MASS IN GROUPS AND CLUSTERS OF GALAXIES}

In order to determine the full stellar mass of a system of
galaxies within the projected radius $R_{200c}$ ($M_K<-21\fm5$)
we used the results of stellar mass data for galaxies from the
SDSS\,DR10 archive which were obtained by fitting
FSPS~\cite{Conroy:Kopylova_n_en} models to $u, g, r, i, z$-band
\mbox {SDSS-photometry} (corrected for extinction). New results on
the photometry of large galaxies (brightest cluster galaxies) are
presented in this release, the luminosity of which has been
underestimated earlier due to excessive background
subtraction~\cite{Bernardi:Kopylova_n_en}.

We should note that not
all galaxies which we included in the sample have spectral data in
SDSS\,DR10 (which includes stellar masses). For example, they are
absent for the brightest galaxy in the A\,2162, AWM\,4, and
A\,1139 systems, the luminosity of which may account for up to
40\% of the total luminosity. Only seven systems (18\% of the
total number) have data available on all galaxies, and the
remaining galaxy systems have no stellar mass measurements (from
3\% to 19\%), and about 30\% for systems RXJ\,1722 and NGC\,5627.
When no stellar mass measurements were available, we estimated
then approximately from the $(u-r)_o$ and $(g-r)_o$ colors, magnitudes,
and the $fracDeV$ parameter.

\begin{figure}[]
\setcaptionmargin{5mm} \onelinecaptionstrue
\includegraphics[scale=0.4,angle=0]{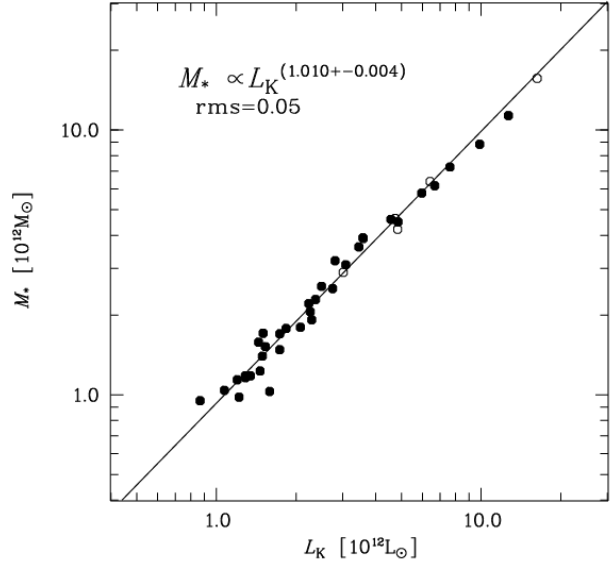}
\captionstyle{normal} \caption{Dependence of the galaxy cluster
stellar mass on $K_s$-luminosity within $R_{200c}$. The line
corresponds to the regression ratio \mbox {$M_*/M_{\odot} \propto
(L_K/L_{\odot})^{1.010\pm0.004}$}. The empty circles show the
galaxy clusters from~\cite{Kopylova0:Kopylova_n_en}.}
\label{M*L:Kopylova_n_en}
\end{figure}

\begin{figure}[]
\setcaptionmargin{5mm} \onelinecaptionstrue
\includegraphics[scale=0.4,angle=0]{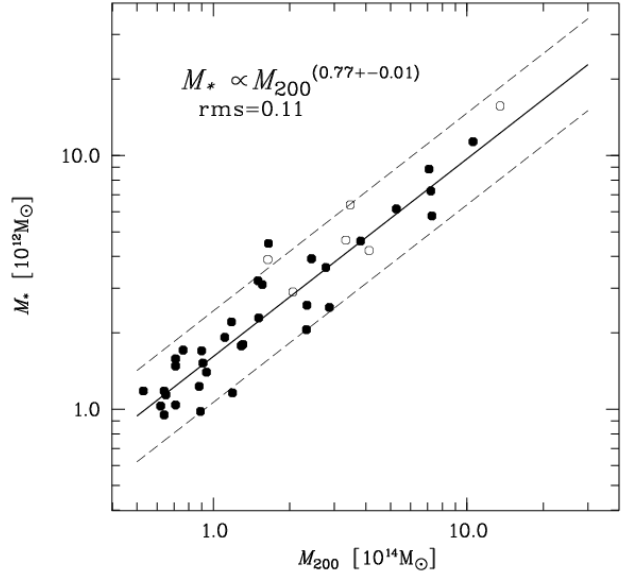}
\captionstyle{normal} \caption{Dependence of the galaxy cluster
stellar mass on the halo mass within $R_{200c}$. The solid line
corresponds to the regression ratio $M_*/M_{\odot} \propto
(M_{200c}/M_{\odot})^{0.77\pm0.004}$, and the dashed lines show
$2\sigma$ deviations from it. The empty circles show the galaxy
clusters from~\cite{Kopylova0:Kopylova_n_en}.}
\label{M*M2:Kopylova_n_en}
\end{figure}

In our earlier
papers~\cite{Kopylova1:Kopylova_n_en,Kopylova2:Kopylova_n_en,Kopylov:Kopylova_n_en}
we used the 2MASX catalog data to determine the
$K_s$-lu\-mi\-no\-si\-ti\-es of the considered galaxy clusters
within $R_{200c}$ (corresponding to \mbox{$M_K<-21\fm0$}). The
mass of the stellar population of the galaxies is best determined
from $K_s$-luminosities. It can be derived by taking into account
the mass--luminosity ratio for spiral and elliptical galaxies. We
compared the derived $K_s$--luminosities with the stellar masses
presented in SDSS\,DR10 and obtained for our sample of 40 galaxy
groups and clusters the following relation:
$$
\begin{array}{rcl}
\log M_{*,200c}& = &(1.010\pm0.004)\\[-5pt]
& \times &  \log L_{K,200c}-(0.144\pm0.050)
\end{array}
$$
\noindent with a standard deviation of
0.05~(Fig.~\ref{M*L:Kopylova_n_en}). Masses and luminosities in
all the expressions are given in $M_{\odot}$ and $L_{\odot}$.
Thus, the stellar mass estimates of the galaxies (and eventually,
of the galaxy clusters) presented in the SDSS\,DR10 catalog agree
well with their near-IR luminosities.
Fig.~\ref{M*M2:Kopylova_n_en} shows the dependence of the stellar
mass of the galaxy systems on their dynamical mass, which we
determined from the radial velocity dispersion. The relation has a
form of
$$
\begin{array}{rcl}
\log M_{*,200c} & = & (0.77\pm0.01)\\[-5pt]
& \times & (\log M_{200c}-14.5)+(12.595\pm0.12)
\end{array}
$$

\noindent with standard deviation 0.11 and has shape similar to
the one we derived earlier for 182 galaxy systems: $L_{K,200c}
\propto M_{200c}^{0.768\pm0.002}$~\cite{Kopylova1:Kopylova_n_en},
and also---for 93 galaxy systems: \mbox{$L_{K,200c} \propto
M_{200c}^{0.72\pm0.04}$}~\cite{Lin:Kopylova_n_en}, and to 55
groups and 6 clusters of galaxies: \mbox{$L_{K,200c} \propto
M_{200c}^{0.64\pm0.06}$}~\cite{Ramella:Kopylova_n_en}.

\begin{figure*}[hbpt]
\setcaptionmargin{5mm} \onelinecaptionstrue
\includegraphics[scale=0.35,angle=0]{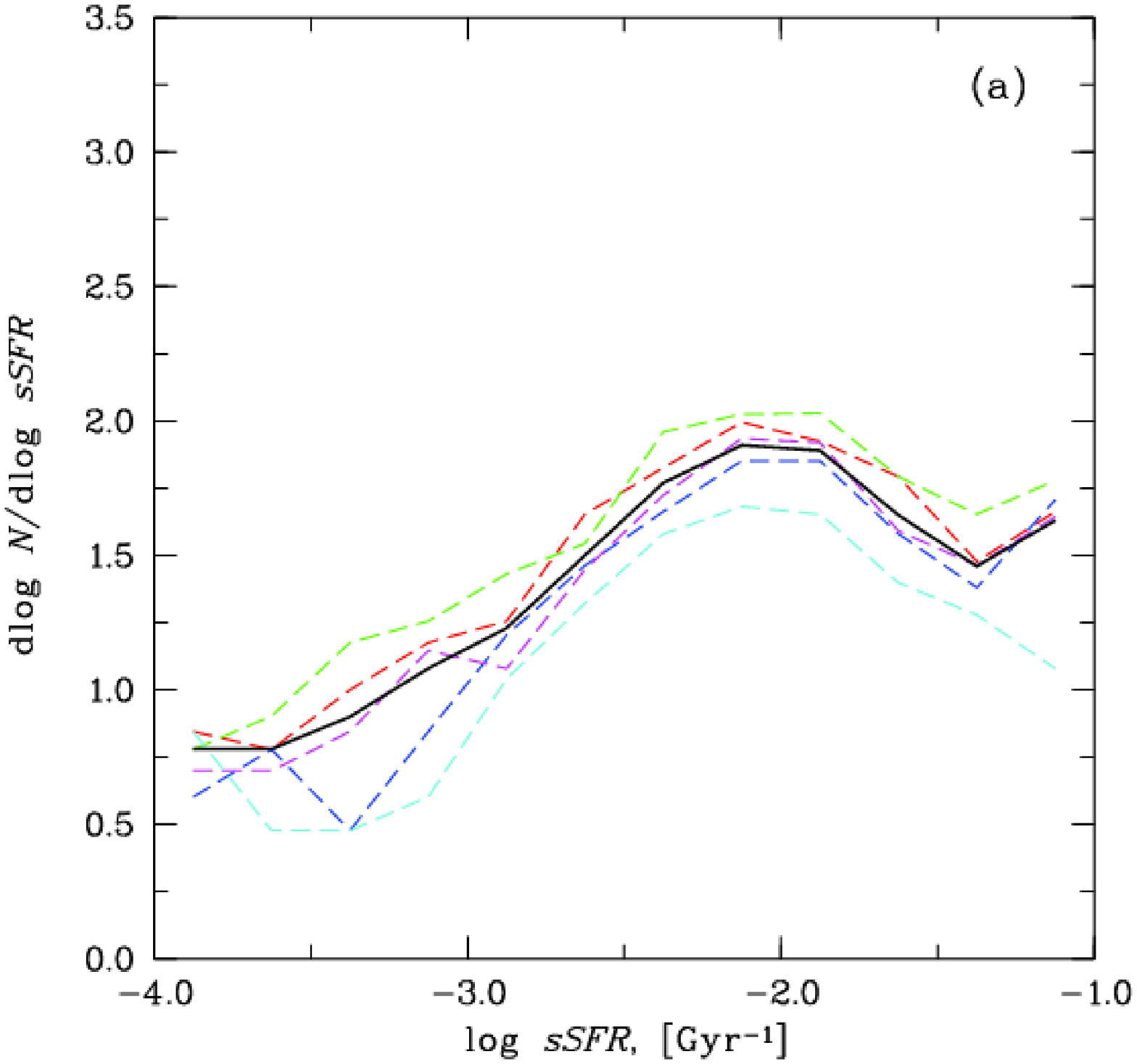}
\includegraphics[scale=0.35,angle=0]{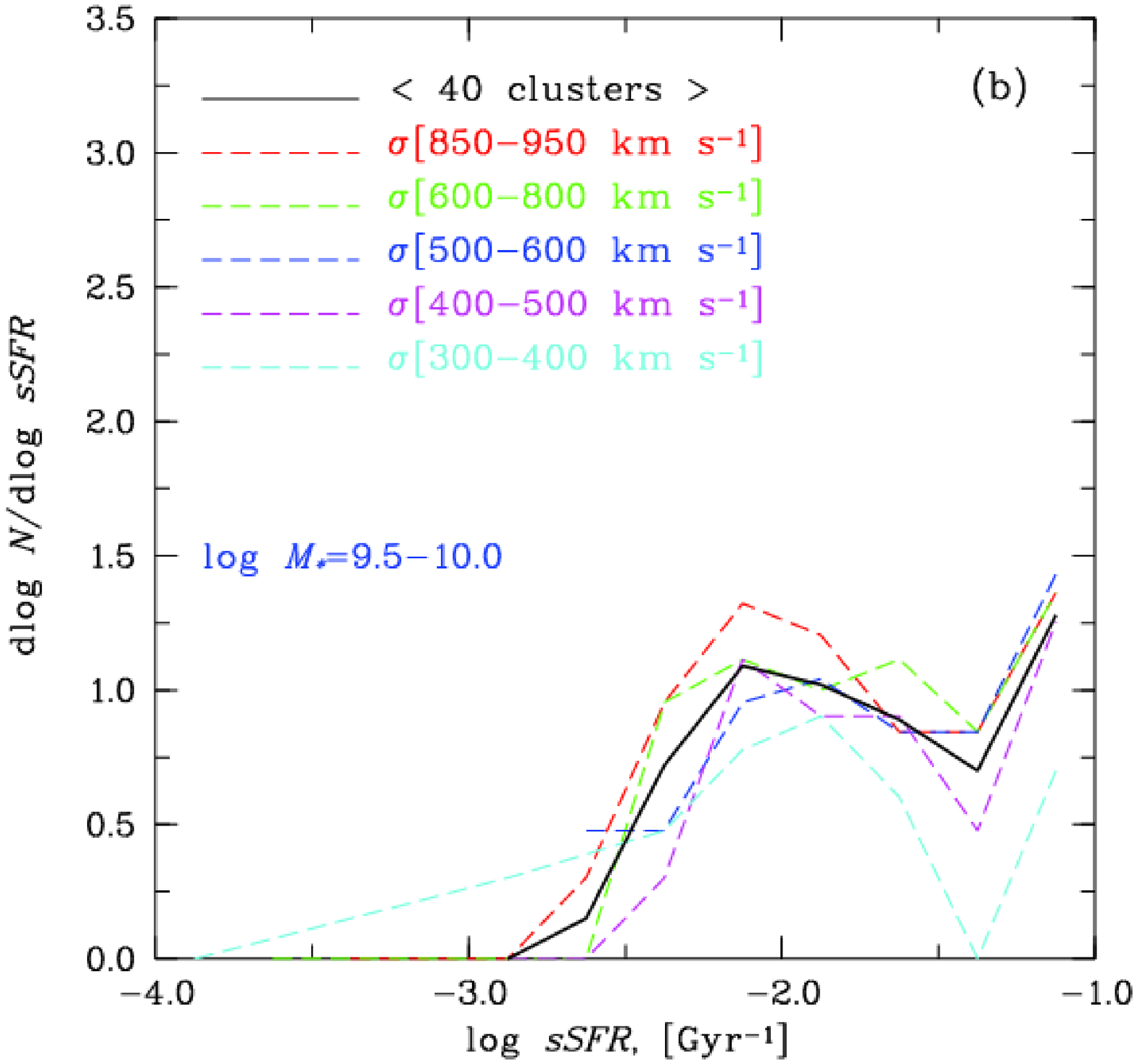}
\includegraphics[scale=0.35,angle=0]{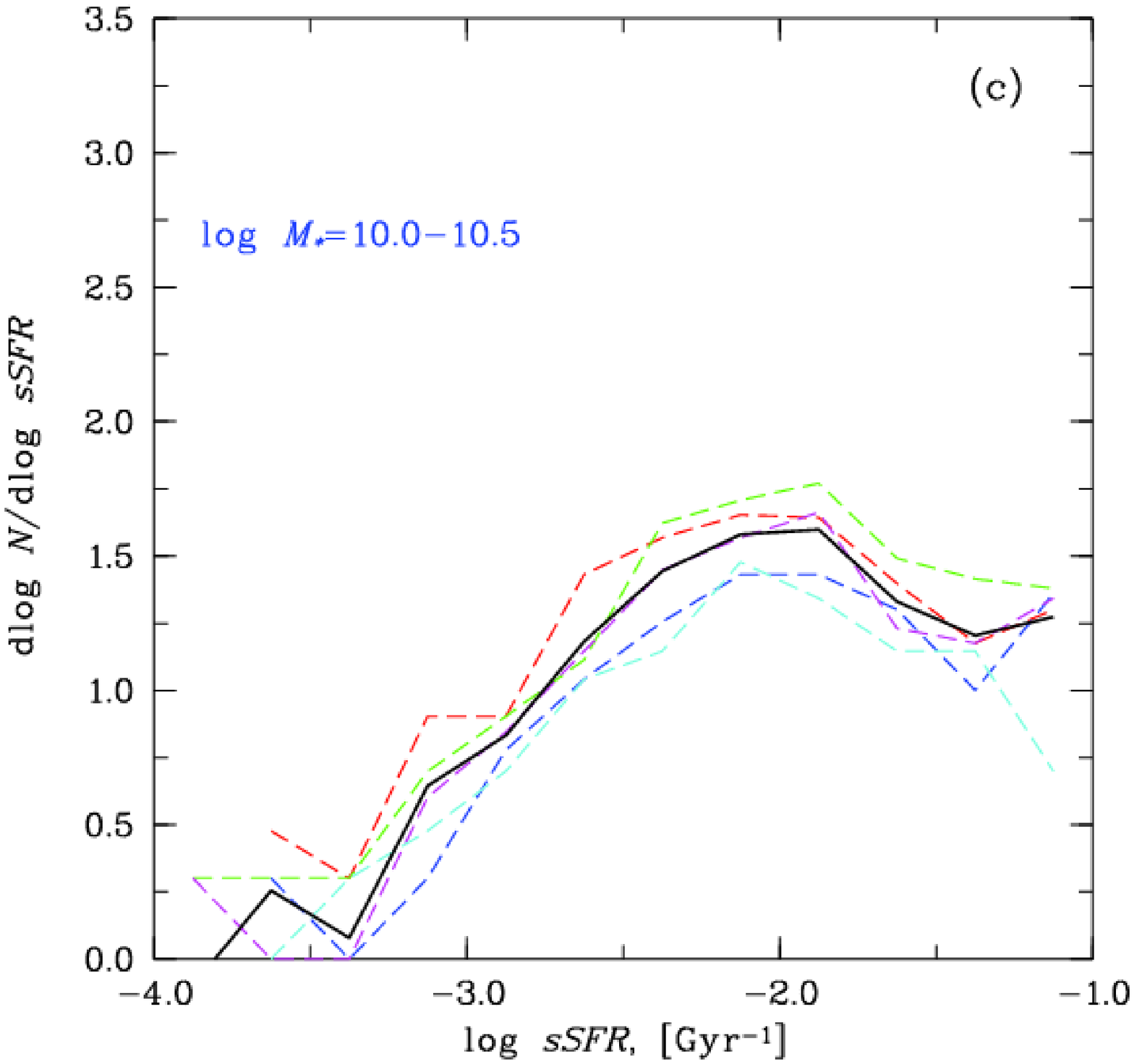}
\includegraphics[scale=0.35,angle=0]{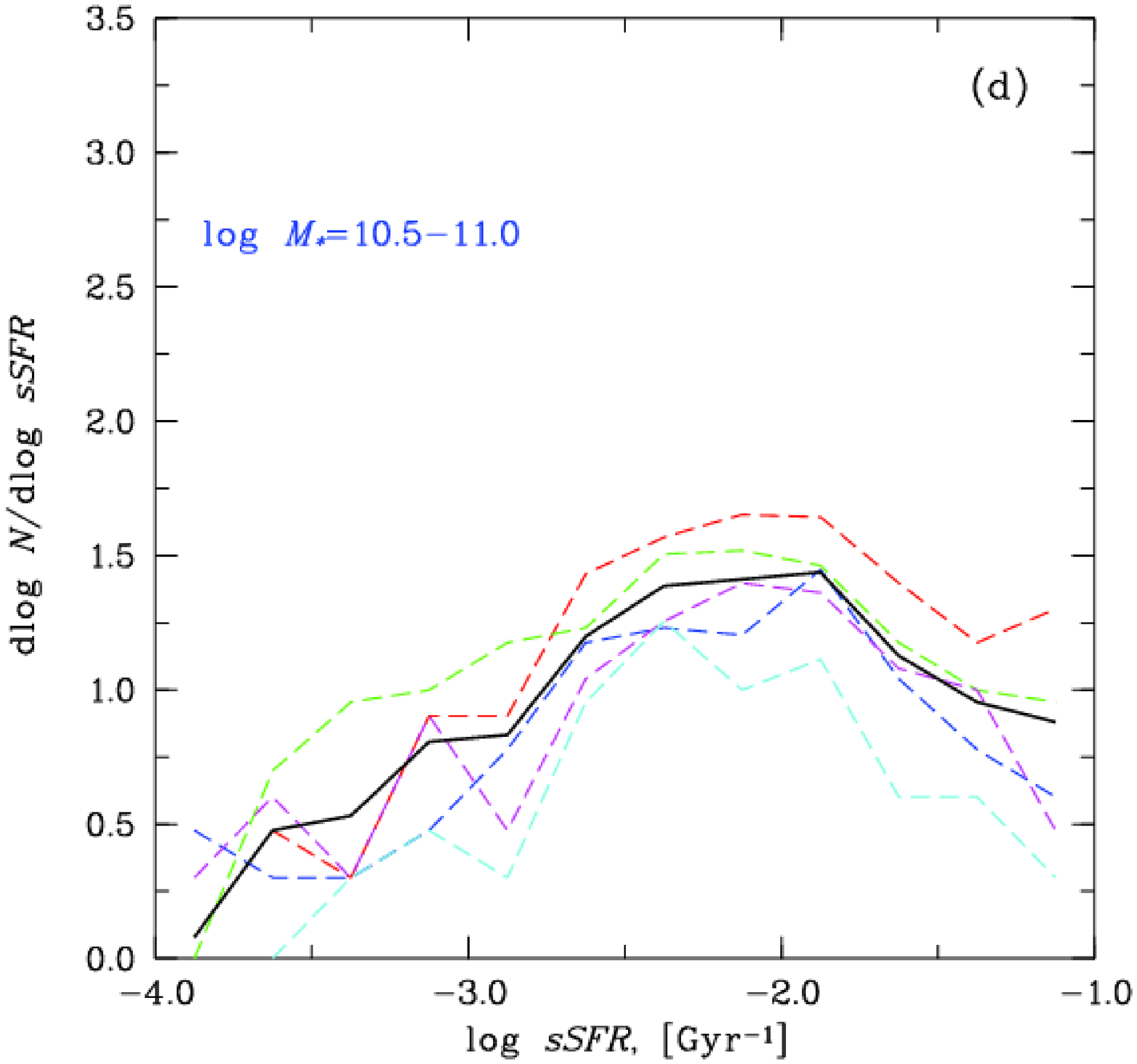}
\includegraphics[scale=0.35,angle=0]{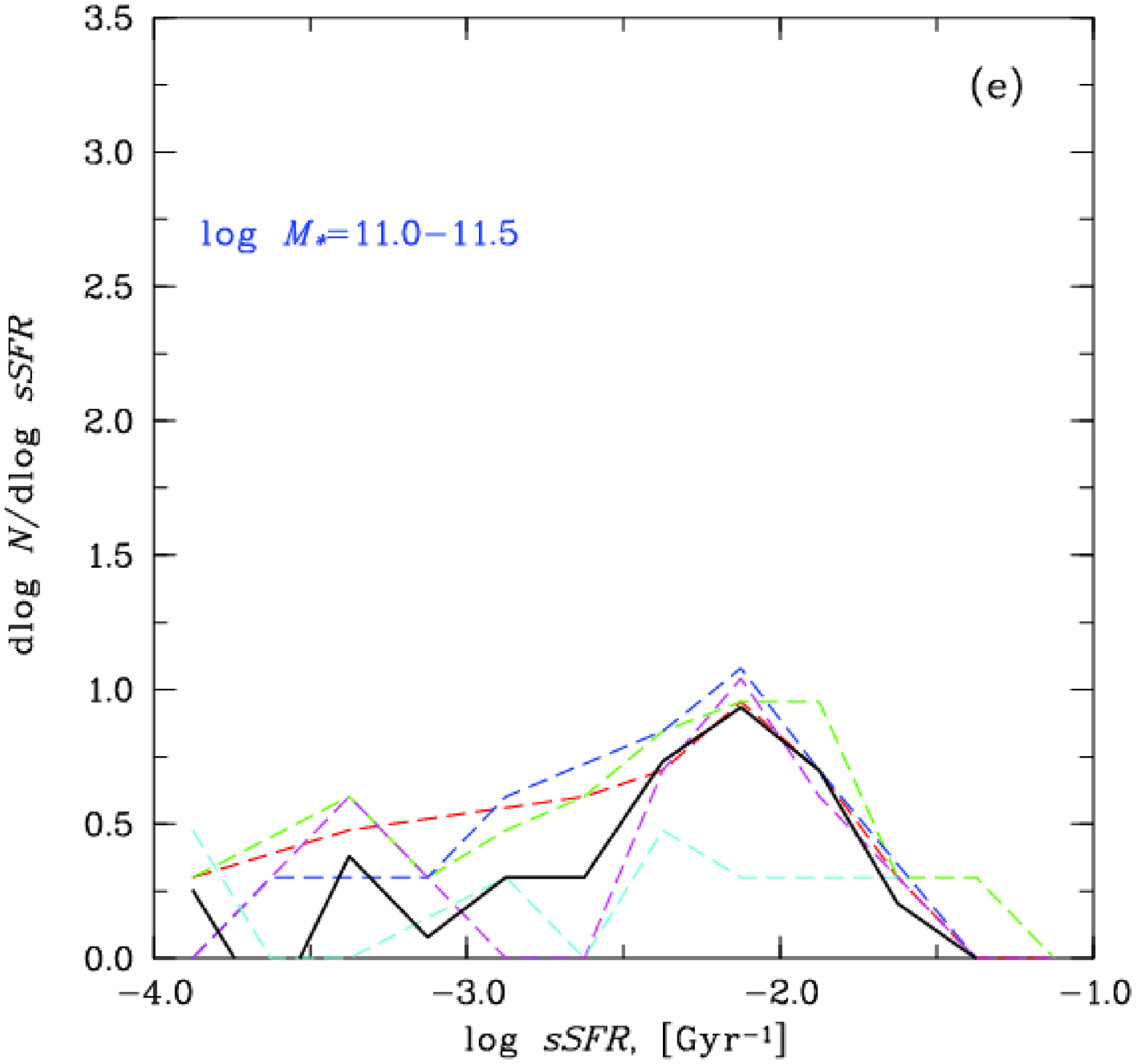}
\captionstyle{normal} \caption{Distribution of galaxies by their
specific star formation rate ($\log sSFR$) for the whole
sample~(a) and with binning by stellar mass~(b)--(e). The solid
line corresponds to the average value, calculated for all
galaxies. Other types of lines show the distributions for groups
and clusters of galaxies with the radial velocity dispersion
$\sigma$ in the range of 850--950, 600--800, 500--600, 400--500
and 300--400~km~s$^{-1}$.} \label{sSFR:Kopylova_n_en}
\end{figure*}

\begin{figure*}[hbpt]
\setcaptionmargin{5mm} \onelinecaptionstrue
\includegraphics[scale=0.35,angle=0]{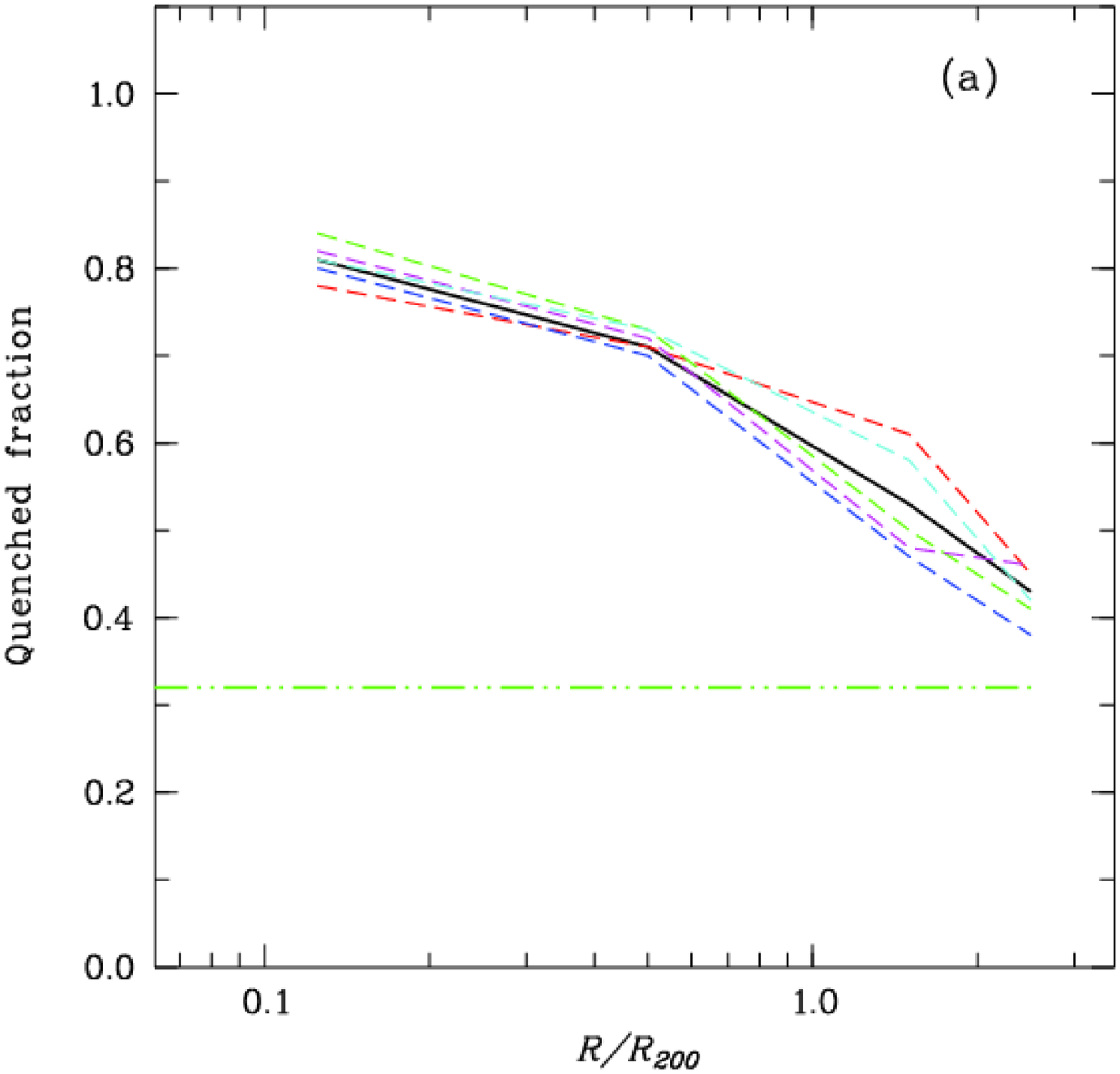}
\includegraphics[scale=0.35,angle=0]{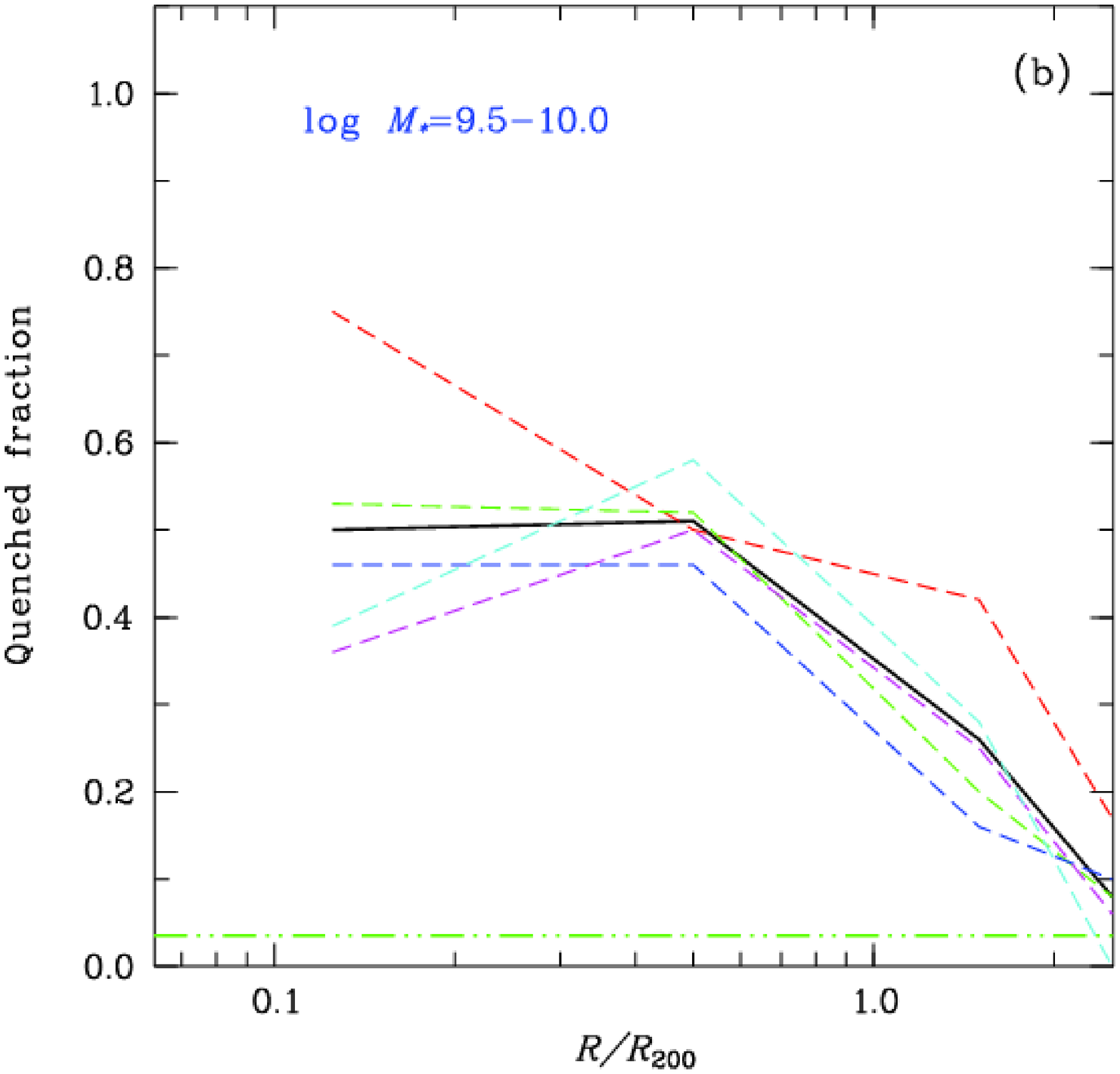}
\includegraphics[scale=0.35,angle=0]{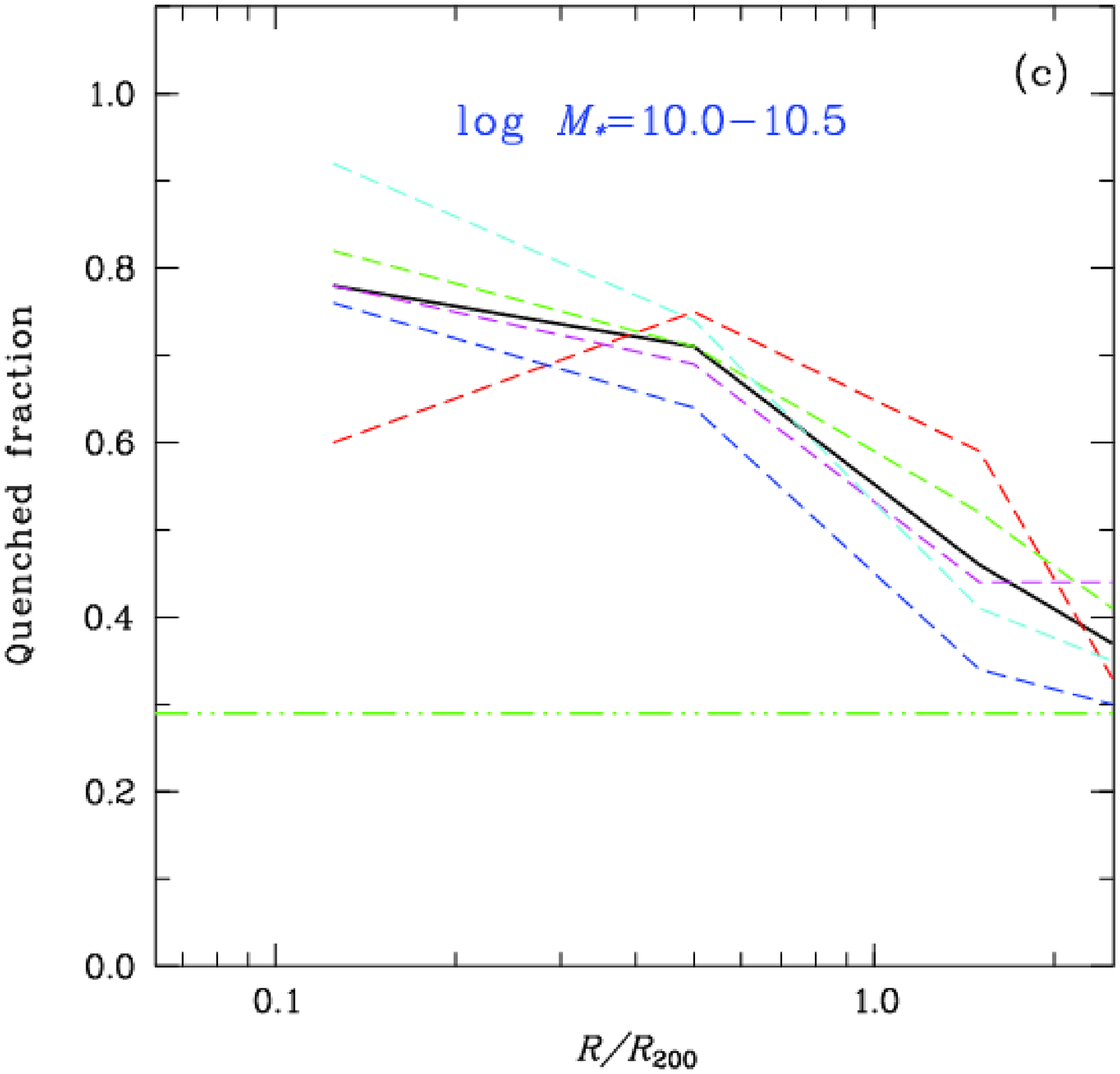}
\includegraphics[scale=0.35,angle=0]{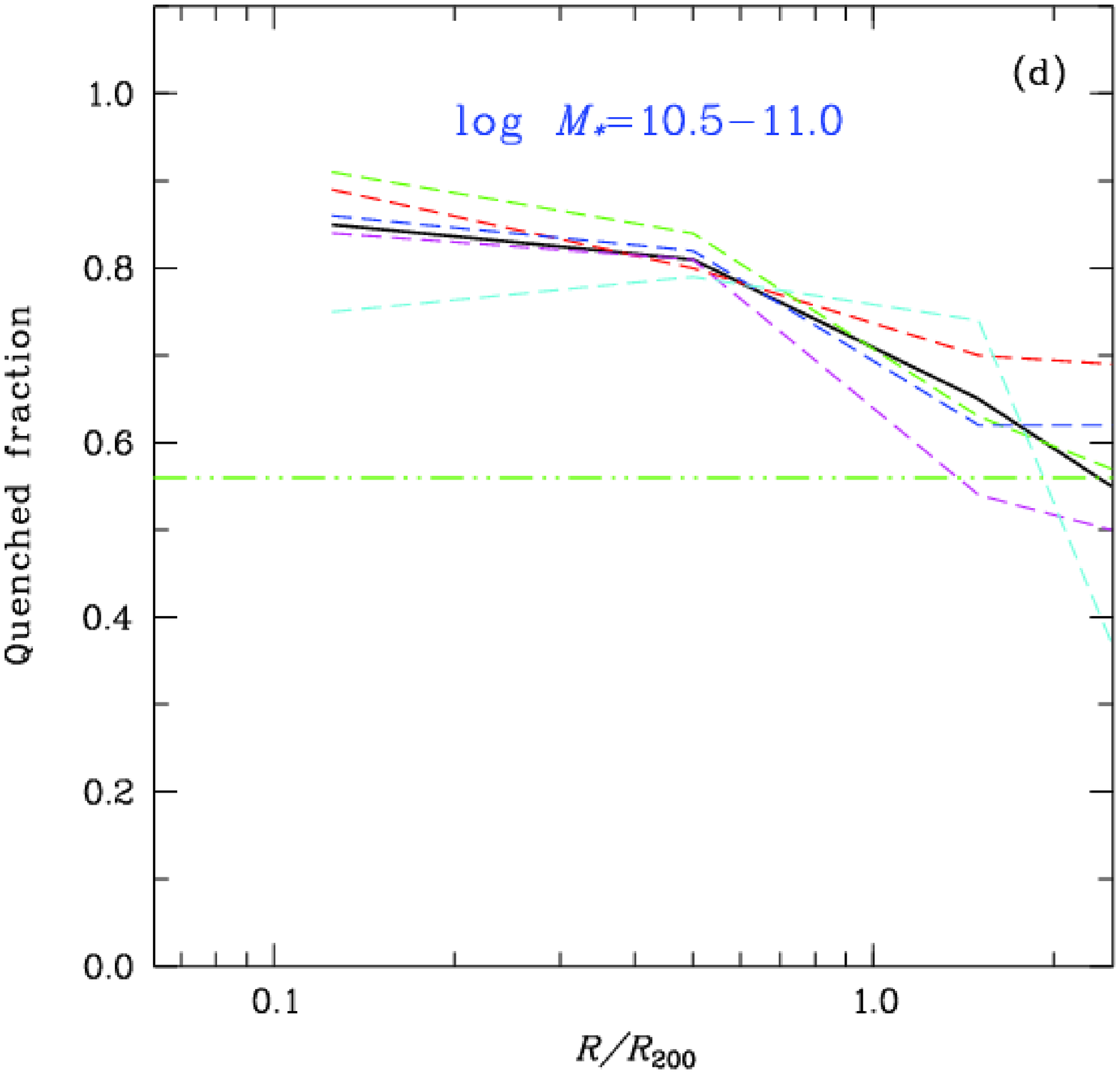}
\includegraphics[scale=0.35,angle=0]{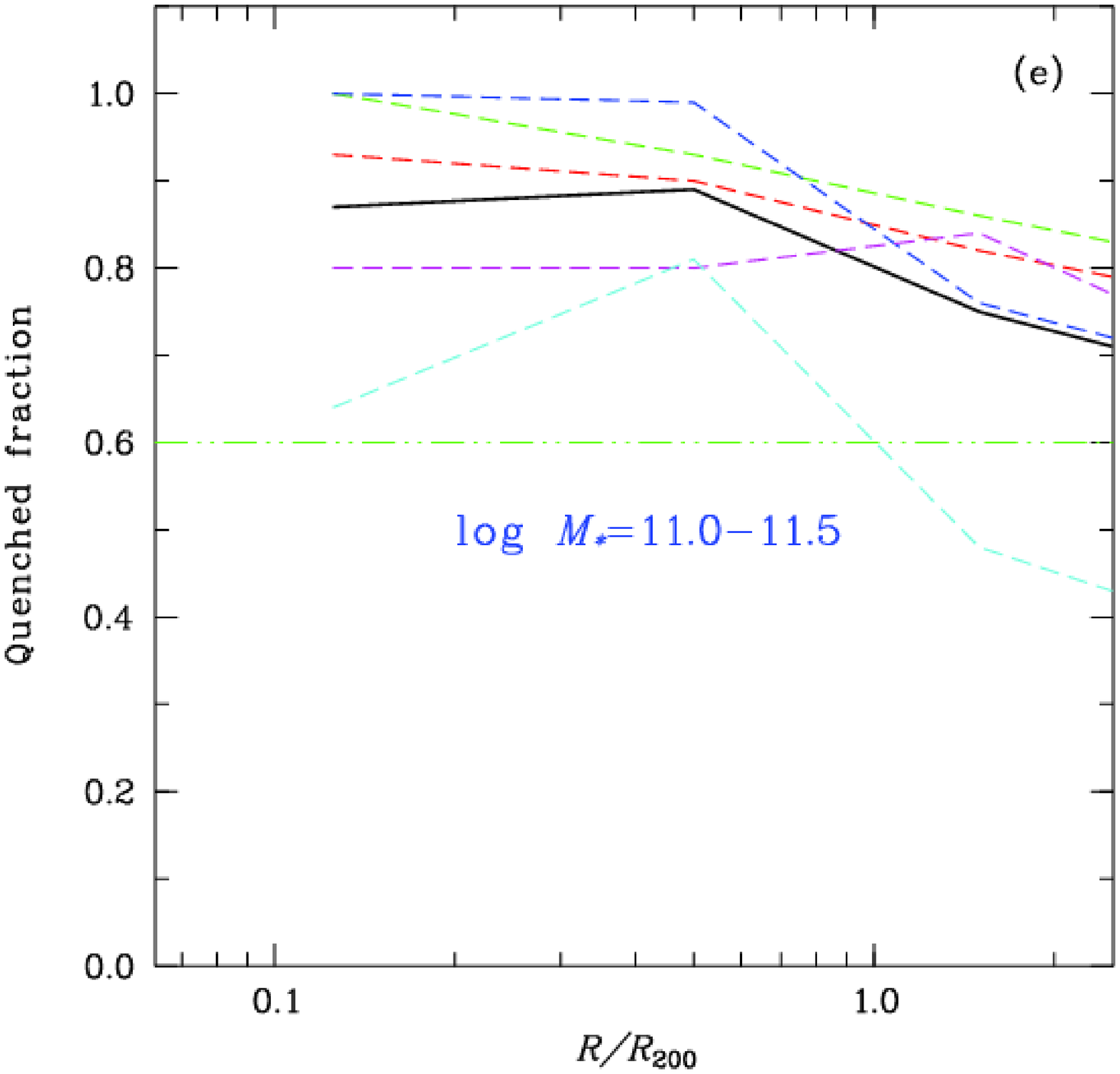}
\captionstyle{normal} \caption{The fraction of galaxies with
quenched star formation ($frac_q$) ($\log sSFR [{\rm Gyr^{-1}}] <
-1.75$) as a function of distance to the cluster center.
The clusters are groups in the same way as in the previous
figures. The solid line corresponds to the average value for all
galaxy systems. Panel~(a) shows $frac_q$, found from all the
galaxies, panels~(b)--(e)---from galaxies binned by stellar mass.
In each panel, the dashed-and-dotted line shows the $frac_q$ in
the field.} \label{FRACQ:Kopylova_n_en}
\end{figure*}

\section{SPECIFIC STAR FORMATION RATE IN GROUPS AND CLUSTERS OF GALAXIES}

The specific star formation rate ($sSFR$) in a galaxy is
determined by the integrated star formation rate divided by its
stellar mass: $sSFR = SFR/M_*$. The SDSS\,DR10 catalog gives the
results of determining the specific star formation rate as well as
the mass of stars in the galaxies. If the conditions $r_{\rm
Pet}<17\fm77$ and \mbox {$\langle \mu_r \rangle <
24\fm5/\Box\arcsec$} are satisfied (the $r$-band Petrosian
magnitude of the galaxy, corrected for galactic extinction, and
the average Petrosian surface brightness corresponding to the
effective radius), then the completeness of SDSS data is estimated
at 99\%~\cite{Strauss:Kopylova_n_en}, and 95\% for bright galaxies. The missing
bright galaxies (mainly early-type with no star formation) within
the radius $R_{200c}$ are taken from the NED database, and also
derived from the color-magnitude diagrams ($u-r, r$, $g-r, r$,
$r-i, r$)~\cite{Kopylova2:Kopylova_n_en}.

The considered galaxy groups and clusters have radial velocity
dispersions $\sigma$ from 300 to 950~km~s$^{-1}$. We divided that
range into the following intervals: 850--950, 600--800, \mbox
{500--600}, 400--500 and 300\mbox{--}400~km~s$^{-1}$ and plotted
the $\log sSFR$ distributions for all the sample
galaxies~(Fig.~\ref{sSFR:Kopylova_n_en}a) and with subsample
binning according to stellar mass: [9.5;~10.0], [10.0; 10.5],
[10.5; 11.0], [11.0; 11.5] $\log
M_*$[$M_{\odot}$]~(Figs.~\ref{sSFR:Kopylova_n_en}b--\ref{sSFR:Kopylova_n_en}e).
Different kinds of lines in Fig.~\ref{sSFR:Kopylova_n_en}
correspond to distributions of galaxies from different radial
velocity dispersion intervals.

A minimum is usually found in the $\log sSFR$ distribution which
separates the active galaxies from quenched galaxies (QG). A clear
bimodality of the distribution and the minimum corresponding to
\mbox {$\log sSFR [{\rm yr^{-1}}] = -11$} are found in, e.g.,
~\cite{Wetzel:Kopylova_n_en} (based on SDSS\,DR7 data). However, the
uncertainties and shifts in SFR measurements for galaxies may
create an illusion of bimodality~\cite{Feldmann:Kopylova_n_en}, and in order to
obtain a single-mode distribution, one must exclude the galaxies
with zero star formation rate.

In our case,  \mbox{$\log sSFR [{\rm Gyr}^{-1}] \sim -1.5$}
or\linebreak \mbox{$\log sSFR [{\rm yr}^{-1}] \sim -10.5$},
correspond to the minimum for the considered clusters
 (Fig.~\ref{sSFR:Kopylova_n_en}a).
We took a common limit for all the systems, equal to \linebreak
\mbox {$\log sSFR [{\rm Gyr^{-1}}] = -1.75$}, the same as in our
earlier paper~\cite{Kopylova0:Kopylova_n_en}. As is evident from
Figs.~\ref{sSFR:Kopylova_n_en}b--\ref{sSFR:Kopylova_n_en}e, the
$\log sSFR$ distribution has a long tail (left) for systems with
the stellar mass in the \mbox{$\log M_*$ = [10.5; 11.5]} interval
(QG region), except for the ranges $\log M_*$ = [10.0; 10.5] and
especially \mbox {$\log M_*$ = [9.5; 10.0]}. The distributions are
similar for all the groups and clusters of galaxies in all stellar
mass ranges, with the exception of the most low-mass groups with
radial velocity dispersions of \mbox {300--400}~km~s$^{-1}$, where
no increase in the number of active galaxies is observed in the
\mbox {$\log M_*$ = [10.0; 10.5]} range
(Fig.~\ref{sSFR:Kopylova_n_en}c). The distribution has a broad
peak \mbox {$\log sSFR = [-2.3; -1.8]$} for all mass ranges except
$\log M_*$ = [11.0; 11.5], where the peak is sharp with the
maximum at \mbox{$\log sSFR \approx -2.2$}, with the exception of
the low-mass group subsample.

\renewcommand\baselinestretch{0.85}
\begin{table*}[bpt!!!]
\setcaptionmargin{0mm} \onelinecaptionstrue \captionstyle{normal}
\caption{Variation of the fraction of quenched galaxies along the
radius} \label{data3:Kopylova_n_en}
\medskip
\begin{tabular}{l|c|c|c|c|c|c}
\hline
\multicolumn{1}{c|}{Cluster} & (0--0.25)$R/R_{200c}$ & (0--1)$R/R_{200c}$ & (1--2)$R/R_{200c}$ &  (2--3)$R/R_{200c}$ & (0--1)$R_{\rm sp}$ & 1$R_{\rm sp}$--3$R/R_{200c}$\\
\hline
\multicolumn{1}{c|}{(1)}&(2)&(3)&(4)&(5)&(6)&(7)\\
\hline
A\,2147     &  $0.79\pm0.36$ &$0.68\pm0.07$  & $0.54\pm0.06$   & $0.48\pm0.05$ & $0.64\pm0.05$ & $0.49\pm0.04$\\
A\,2063     &  $0.86\pm0.18$ &$0.71\pm0.10$  & $0.40\pm0.10$   & $0.55\pm0.10$ & $0.66\pm0.09$ & $0.50\pm0.08$\\
A\,1367     &  $0.74\pm0.17$ &$0.68\pm0.09$  & $0.58\pm0.08$   & $0.39\pm0.10$ & $0.65\pm0.08$ & $0.46\pm0.09$\\
A\,2199     &  $0.88\pm0.20$ &$0.72\pm0.08$  & $0.60\pm0.08$   & $0.43\pm0.07$ & $0.66\pm0.06$ & $0.43\pm0.07$\\
A\,1185     &  $0.68\pm0.18$ &$0.56\pm0.08$  & $0.50\pm0.09$   & $0.44\pm0.10$ & $0.55\pm0.07$ & $0.43\pm0.08$\\
MKW\,03s    &  $1.00\pm0.30$ &$0.81\pm0.12$  & $0.56\pm0.15$   & $0.38\pm0.15$ & $0.78\pm0.11$ & $0.40\pm0.12$\\
NGC\,6338   &  $0.84\pm0.29$ &$0.65\pm0.14$  & $0.52\pm0.18$   & $0.60\pm0.16$ & $0.61\pm0.12$ & $0.60\pm0.15$\\
NGC\,6107   &  $0.80\pm0.27$ &$0.83\pm0.15$  & $0.61\pm0.18$   & $0.34\pm0.11$ & $0.75\pm0.12$ & $0.40\pm0.10$\\
RXC\,J1722  &  $0.86\pm0.33$ &$0.63\pm0.15$  & $0.41\pm0.18$   & $0.33\pm0.17$ & $0.62\pm0.14$ & $0.33\pm0.14$\\
MKW\,04     &  $0.87\pm0.33$ &$0.85\pm0.20$  & $0.25\pm0.28$   & $0.27\pm0.18$ & $0.81\pm0.20$ & $0.25\pm0.16$\\
UGC\,04991  &  $0.50\pm0.22$ &$0.56\pm0.14$  & $0.45\pm0.24$   & $0.20\pm0.13$ & $0.59\pm0.14$ & $0.22\pm0.11$\\
A\,1983     &  $0.80\pm0.24$ &$0.78\pm0.13$  & $0.49\pm0.12$   & $0.43\pm0.17$ & $0.71\pm0.10$ & $0.42\pm0.12$\\
MKW\,08     &  $0.92\pm0.38$ &$0.89\pm0.13$  & $0.54\pm0.13$   & $0.55\pm0.16$ & $0.78\pm0.13$ & $0.56\pm0.14$\\
NGC\,5098   &  $0.73\pm0.29$ &$0.72\pm0.16$  & $0.32\pm0.11$   & $0.50\pm0.16$ & $0.60\pm0.12$ & $0.42\pm0.12$\\
RBS\,858    &  $1.00\pm0.39$ &$0.79\pm0.18$  & $0.87\pm0.33$   & $0.25\pm0.20$ & $0.79\pm0.16$ & $0.57\pm0.25$\\
NGC\,2795   &  $0.54\pm0.28$ &$0.52\pm0.16$  & $0.50\pm0.23$   & $0.57\pm0.21$ & $0.53\pm0.15$ & $0.53\pm0.16$\\
MKW\,04s    &  $1.00\pm0.63$ &$0.73\pm0.20$  & $0.62\pm0.36$   & $0.50\pm0.43$ & $0.70\pm0.18$ & $0.60\pm0.44$\\
VV\,196     &  $0.83\pm0.36$ &$0.56\pm0.19$  & $0.12\pm0.09$   & $0.57\pm0.25$ & $0.52\pm0.16$ & $0.35\pm0.13$\\
RXC\,J1033  &  $0.83\pm0.38$ &$0.71\pm0.17$  & $0.67\pm0.20$   & $0.33\pm0.17$ & $0.70\pm0.14$ & $0.40\pm0.17$\\
AWM\,1      &  $0.75\pm0.40$ &$0.73\pm0.20$  & $0.33\pm0.17$   & $0.67\pm0.30$ & $0.72\pm0.20$ & $0.48\pm0.17$\\
AWM\,4      &  $1.00\pm0.53$ &$0.83\pm0.24$  & $0.44\pm0.25$   & $0.33\pm0.22$ & $0.82\pm0.23$ & $0.28\pm0.16$\\
NGC\,7237   &  $1.00\pm0.45$ &$0.81\pm0.20$  & $0.71\pm0.30$   & $0.50\pm0.43$ & $0.80\pm0.17$ & $0.40\pm0.33$\\
NGC\,3158   &  $0.77\pm0.32$ &$0.68\pm0.23$  & $0.62\pm0.36$   & --            & $0.69\pm0.21$ & $0.60\pm0.44$\\
RXC\,J1511  &  $0.86\pm0.48$ &$0.75\pm0.23$  & $0.46\pm0.24$   & $0.46\pm0.24$ & $0.71\pm0.21$ & $0.44\pm0.19$\\
NGC\,5171   &  $0.54\pm0.25$ &$0.61\pm0.17$  & $0.56\pm0.23$   & $0.30\pm0.20$ & $0.61\pm0.14$ & $0.27\pm0.18$\\
NGC\,3119   &  $0.78\pm0.39$ &$0.73\pm0.22$  & $0.26\pm0.13$   & $0.20\pm0.22$ & $0.62\pm0.18$ & $0.25\pm0.14$\\
A\,1228B    &  $0.93\pm0.36$ &$0.80\pm0.24$  & $0.56\pm0.22$   & $0.50\pm0.27$ & $0.74\pm0.18$ & $0.47\pm0.21$\\
A\,2162     &  $0.88\pm0.45$ &$0.77\pm0.25$  & $0.71\pm0.30$   & $0.50\pm0.31$ & $0.83\pm0.23$ & $0.43\pm0.21$\\
NGC\,2783   &  $0.50\pm0.31$ &$0.56\pm0.23$  & $0.60\pm0.44$   & $1.00\pm0.63$ & $0.56\pm0.22$ & $0.88\pm0.45$\\
A\,1177     &  $0.75\pm0.40$ &$0.86\pm0.28$  & $0.80\pm0.38$   & $0.50\pm0.43$ & $0.86\pm0.24$ & $0.57\pm0.36$\\
NGC\,6098   &  $1.00\pm0.82$ &$0.50\pm0.27$  & $0.20\pm0.22$   & --            & $0.36\pm0.19$ & $0.50\pm0.61$\\
UGC\,07115  &  $0.67\pm0.43$ &$0.77\pm0.21$  & $0.71\pm0.42$   & $0.50\pm0.31$ & $0.75\pm0.19$ & $0.50\pm0.27$\\
NGC\,2832   &  $0.88\pm0.45$ &$0.88\pm0.26$  & $0.85\pm0.28$   & $0.90\pm0.41$ & $0.85\pm0.21$ & $0.94\pm0.34$\\
NGC\,5627   &  $0.86\pm0.48$ &$0.74\pm0.26$  & $0.60\pm0.31$   & $0.17\pm0.18$ & $0.77\pm0.24$ & $0.20\pm0.15$\\
\hline
Total sample ($N=40$)  &  $0.81\pm0.02$ &$0.71\pm0.02$  & $0.53\pm0.02$   & $0.43\pm0.02$ & $0.68\pm0.02$ & $0.44\pm0.02$\\
\hline
\end{tabular}
\end{table*}
\renewcommand\baselinestretch{1.0}

In galaxy clusters the density decreases with increasing distance
from the selected center, and the position of a galaxy at a
certain radius is related to the time of its entering the cluster
(see, e.g.,~\cite{Haines:Kopylova_n_en}).
Fig.~\ref{FRACQ:Kopylova_n_en}  shows the fraction of quenched
galaxies QG as a function of projected cluster radius normalized
to radius $R_{200c}$. Fig.~\ref{FRACQ:Kopylova_n_en}a shows the
total fraction of QG without separating by stellar mass, and
Figs.~\ref{FRACQ:Kopylova_n_en}b\mbox{--}\ref{FRACQ:Kopylova_n_en}e
show the QG for each stellar mass range individually. Different
lines show the variations of the fraction of QG, corresponding to
the radial velocity dispersion bin. The total fraction of QG
varies from the average value $0.81\pm0.02$ at the
distance\linebreak $(0-0.25)R/R_{200c}$ to $0.43\pm0.02$ at
$(2-3)R/R_{200c}$ (or $0.44\pm0.02$ beyond the radius $R_{\rm
sp}$). Thus, the fraction of QG decreases at $3R/R_{200c}$ by
almost 50\%. We found the average (for two fields) fraction of
galaxies with quenched star formation, for which the condition
\mbox {$\log sSFR [{\rm Gyr^{-1}}] < -1.75$} is met. It is equal
to $0.32\pm0.07$, which is 60\% less than that in the central
regions of the galaxy clusters, or 53\% smaller than within the
$R_{\rm sp}$ radius, or 26\% smaller than within
(2--3)$R/R_{200c}$. As is evident from
Fig.~\ref{FRACQ:Kopylova_n_en}, all $\sigma$ bins (essentially,
bins by mass) show the same behavior along the radius for fixed
$M_*$, except the \mbox {$\sigma$ = (300--400)}~km~s$^{-1}$ bin
(Fig.~\ref{FRACQ:Kopylova_n_en}e). For stellar masses in the $\log
M_*$ = [10.5; 11.0] range (Fig.~\ref{FRACQ:Kopylova_n_en}d), the
fraction of QG in the galaxy system outskirts are close to the
field values. For the remaining stellar masses
(Figs.~\ref{FRACQ:Kopylova_n_en}b,~\ref{FRACQ:Kopylova_n_en}c
and~\ref{FRACQ:Kopylova_n_en}e) the QG fraction is higher, on
average, than that in the field.

In~\cite{Haines:Kopylova_n_en} it is shown for 30 galaxy clusters with
$0.15<z<0.30$ that the fraction of active galaxies is lower than
that in the field even within the radius $3R_{200c}$, i.e., the
fraction of QG is higher. According to Wetzel et
al.~\cite{Wetzel1:Kopylova_n_en}, an excess of quenched galaxies is observed
beyond the virial radius of the galaxy systems (up to $2.5R_{\rm
vir}$), based on SDSS\,DR7 data. Thus, the fraction of QG remains
higher than that in the field even beyond the virialized regions.

Table~\ref{data3:Kopylova_n_en} shows the results of measuring the QG fraction
along the normalized cluster radius. The first column in the table
gives the system name, the remaining columns give the variation
intervals for the radii ($R/R_{200c}$ and $R_{\rm sp}$).

\section{CONCLUSIONS}

The observational data show that galaxies in groups and clusters
differ from field galaxies. The existing correlation between the
morphology of the galaxies and the density of their outskirts
region becomes the morphology---radius ratio in galaxy clusters,
since the density decreases with increasing radius. In this work,
we investigated the central regions and nearest outskirts of
galaxy systems (up to $3R_{200c}$ in projection). For the
considered galaxy cluster sample we determined: the specific star
formation rate, the quenched galaxy fraction, the fraction of
early type galaxies on the ``red sequence'' in comparison with the
field data. We also found for each galaxy system the stellar mass
within the $R_{200c}$ radius and compared it with the halo mass,
measured from the radial velocity dispersion. The characteristics
of 34 galaxy systems from the studied sample are presented in this
paper, and the results for six systems have been published
earlier~\cite{Kopylova0:Kopylova_n_en}.

We obtained the following results for all 40 systems:

\begin{list}{}{
\setlength\leftmargin{5mm} \setlength\topsep{2mm}
\setlength\parsep{0mm} \setlength\itemsep{2mm} }

\item 1. For the considered galaxy systems \linebreak \mbox
{($\log M_{200c}/M_{\odot} = [13.7; 15.00]$)} the galaxy stellar
mass within the radius $R_{200c}$, determined from SDSS\,DR10
data, corresponds to the IR\mbox{-}lu\-mi\-no\-si\-ty measured
from the 2MASX data:\linebreak $M_{*,200c}/M_{\odot} \propto
(L_{K,200c}/L_{\odot})^{1.010}$. The dependence between the
stellar mass and the dynamical mass (halo) of the system has the
form of
$$
\begin{array}{rcl}
\log M_{*,200c}M_{\odot} & = & (0.77\pm0.01)\\[-5pt]
& \times & (\log M_{200c}M_{\odot}-14.5)+(12.60\pm0.12)
\end{array}
$$
\item 2. A radial gradient of $sSFR$ is present in the galaxy
clusters. The fraction of QG is maximal ($\log sSFR [\rm
{Gyr^{-1}}]<-1.75$) in the central regions and, based on the
average for all groups and clusters, amounts to $0.81\pm0.02$
beyond $R_{\rm sp}$ (within $(2-3)R/R_{200c}$)---$0.44\pm0.03$.
This value, obtained for the outskirts of the galaxy clusters is
higher than that in the field by about 27\%. The highest
variations of the quenched galaxy fraction with radius are
observed in galaxy systems with stellar masses of $\log
M_*/M_{\odot} = 9.5$--$10.0$.

\item 3. The ``red sequence'' of early type galaxies with small
scatter is the main property of galaxy clusters, since bright
early type galaxies constitute 60--70\% of their virialized
regions. We show that $frac_E$ beyond $R_{\rm sp}$ (up to
$(2-3)R/R_{200c}$) becomes the same as that in the field.
\end{list}

\begin{acknowledgments}
This research has made use of the NASA/IPAC Extragalactic Database
(NED, \url{http://nedwww.}\linebreak\url{.ipac.caltech.edu}),
which is operated by the Jet Propulsion Laboratory, California
Institute of Technology, under contract with the National
Aeronautics and Space Administration, Sloan Digital Sky Survey
(SDSS, \url{http://www.sdss.org}), which is supported by Alfred P.
Sloan Foundation, the participant institutes of the SDSS
collaboration, National Science Foundation, and the United States
Department of Energy and Two Micron All Sky Survey (2MASS,
\url{http://www.ipac.}\linebreak\url{.caltech.edu/2mass/releases/allsky/}).
\end{acknowledgments}


\section*{CONFLICT OF INTEREST}
The authors declare no conflict of interest.

\begin{center}
\refname
\end{center}

\end{document}